\documentclass[10pt,twocolumn]{IEEEtran}

\usepackage{adjustbox}

\usepackage{diagbox}
\usepackage{enumitem}
\usepackage{amsmath,graphics,amssymb,epsfig,subfigure,color,cite}
\usepackage{array}
\usepackage{multirow}
\usepackage{enumerate}
\usepackage{algorithmic}
\usepackage{algorithm}
\usepackage{bm}
\usepackage{float}
\usepackage{makecell}
\usepackage{lipsum}
\usepackage{capt-of}
\usepackage{amsmath}
\usepackage{amssymb}

\floatname{algorithm}{Algorithm}

\begin{document}
        \title{\fontsize{22}{28}\selectfont Geometric Cross-Modal Token Selection for Latency-Constrained Multimodal Token Communication}
	 \author{Joohyuk Park, Junyong Shin, Yongjeong Oh, Jihong Park, and Yo-Seb Jeon
    \thanks{Joohyuk Park, Junyong Shin, and Yo-Seb Jeon are with the Department of Electrical Engineering, POSTECH, Pohang, Gyeongbuk 37673, Republic of Korea (e-mail: joohyuk.park@postech.ac.kr; sjyong@postech.ac.kr;
    yoseb.jeon@postech.ac.kr). }
    \thanks{Yongjeong Oh, Jihong Park are with the Singapore University of Technology and Design (SUTD), Singapore 487372 (e-mail: yongjeong\_oh@sutd.edu.sg;
    jihong\_park@sutd.edu.sg).}
    \thanks{Yo-Seb Jeon and Jihong Park are corresponding authors.}
    }  
	\vspace{-2mm}
	
	\maketitle
 
	\vspace{-12mm}

    \begin{abstract}
        This paper proposes a geometry-based joint cross-modal token selection framework for latency-constrained multimodal token communications.
        To capture cross-modal token dependencies, we leverage the cross-attention mechanism to project modality-specific tokens into a shared query-key space, where the modality with the fewest tokens serves as the anchor modality and the others as non-anchor modalities. Inspired by germ-grain models, we define an angular-distance metric and construct semantic grain regions around anchor queries. Based on this geometric representation, we identify cross-modal evidence shared across multiple anchor queries in this space and develop an intersection-based token selection (IBS) strategy that prioritizes non-anchor tokens whose key are covered by multiple grain regions. We further develop an erasure-aware extension, termed robust-IBS (R-IBS), for token-wise erasure channels using an expected angular-distance formulation. In both IBS and R-IBS, the grain regions are optimized for individual queries under a latency constraint, using block coordinate descent and a low-complexity greedy algorithm.  
        Simulations corroborate the effectiveness of IBS and R-IBS under latency-constrained and token-wise erasure channels, achieving up to $31.6$\% and $29.2$\% task accuracy gains on visual question answering (VQA) and audio-visual question answering (AVQA) tasks, respectively, over existing token selection baselines.
    \end{abstract}


    \section{Introduction}\label{Sec:Intro}
        Recent advancements in wireless networks have shifted the focus of communication from accurate bit reconstruction to the effective delivery of task-relevant information, known as task-oriented or semantic communication \cite{SC_1}. This paradigm has become increasingly important with the rapid development of multimodal large language models (MLLMs) \cite{MLLM_ref1}, including vision-language models (VLMs) \cite{vlm_ref2}, which perform inference by jointly reasoning over information from multiple modalities. A representative example is visual question answering (VQA), where an image and a text query are jointly processed to generate an answer. 

        In wireless MLLM applications, a mobile user transmits multimodal inputs to a remote receiver, such as an edge server, hosting a Transformer-based MLLM. Transformer architectures represent these inputs as sequences of high-dimensional embedding tokens \cite{mqt}, each corresponding to a modality-specific semantic unit, such as an image patch, a text subword, or an audio segment. This setting gives rise to multimodal token communication, where tokens from different modalities are delivered over wireless channels for remote multimodal inference.

        However, limited transmit power and bandwidth constrain the achievable transmission rate, while the communication payload grows with both the number of tokens and their embedding dimensions. Transmitting all tokens over a rate-limited wireless link may therefore violate the target latency constraint. This motivates selective token transmission, in which the transmitter prioritizes the tokens most relevant to the downstream inference task. Since the receiver-side MLLM jointly reasons over multiple modalities, token relevance may depend not only on individual token content but also on semantic relationships across modalities.

        To address this problem, we propose a geometry-based framework for jointly selecting cross-modal tokens under a target latency constraint. Here, cross-modal tokens refer to token representations updated through cross-attention by incorporating information from other modalities. The proposed framework exploits the geometric relationships among these tokens to identify shared semantic information across modalities. The key challenge is to select the most informative cross-modal tokens for transmission, thereby maximizing multimodal inference performance within the latency budget.



    \subsection{Literature Review}
        \subsubsection{Token Compression} 
        Token compression and selection techniques have been widely studied to reduce the computational and memory costs of Transformer-based multimodal models. In VLMs, these methods are typically applied after the projector module, where visual features are mapped into the language-aligned embedding space \cite{Chu_MobileVLMv2_2024, Cha_Honeybee_2024, qformer, qwenvl, tg_llava}. Existing approaches can be broadly categorized into transformation-based and query-based methods. Transformation-based methods, such as pooling or convolutional downsampling, compress token sequences through fixed operations, but may blur fine-grained spatial details \cite{Chu_MobileVLMv2_2024, Cha_Honeybee_2024}. Query-based methods employ learnable latent queries to aggregate information from visual tokens, thereby reducing the number of tokens used by the downstream model \cite{qformer, qwenvl, tg_llava}. Despite their effectiveness for model-side compression, these methods are not designed for wireless token transmission, as they typically assume reliable delivery and fixed compression rates. Moreover, existing studies largely focus on visual-token compression, rather than joint token compression or selection across multiple modalities under dynamic wireless constraints.

    \subsubsection{Token Communication}
        Recently, token-based communication frameworks have emerged in the wireless communication domain, where transformer tokens serve as the basic unit of information transmission \cite{Wei_TokenFramework_2026, Li_TaskOriented_2026, Li_ToDMA_2026}. 
        To reduce transmission latency, several token selection and pruning strategies have been proposed. 
        In single-modal settings, existing methods exploit token importance or contextual predictability to remove less informative tokens \cite{Devoto_Adaptive_2026, Shin_ContextAware_2026}. In multimodal settings, recent studies further use cross-modal cues, such as text-guided image token reconstruction or textual intent-guided video token communication, to recover or prioritize task-relevant tokens \cite{Liu_TextGuided_2025, Men_VideoTokenCom_2026}.
        Despite these efforts, they often rely on fixed compression ratios, learned compression modules, or signal-to-noise ratio (SNR)-specific training, limiting adaptability to time-varying channels \cite{Wei_TokenComm_2025, Hao_TokenCom_2026, Ying_JSCCM_2026}.
        Moreover, most existing methods lack a principled mechanism for joint token selection based on fine-grained cross-modal alignment. Although multimodal foundation models and early semantic communication frameworks often align heterogeneous modalities through contrastive learning \cite{vlm_ref1, Cicchetti_TRIANGLE_2025}, such alignment is typically performed at the sample or global representation level. For multimodal token communication, however, it is necessary to identify which tokens from different modalities are mutually related and should be jointly preserved under latency constraints.

    \subsubsection{Token-level Cross-Modal Alignment for Token Selection}
        Following global-level alignment, multimodal learning studies have increasingly investigated token-level cross-modal alignment to capture finer semantic correspondences across modalities. Such alignment has been used for text-guided reconstruction \cite{Liu_TextGuided_2025}, VLM-based feature learning \cite{Hao_TokenCom_2026}, and cross-modal representation learning \cite{wang2023ca_}. From a communication perspective, recent information-theoretic studies have further highlighted the potential of mapping multimodal sources into a unified semantic embedding space to improve communication efficiency \cite{Unifying_Modalities_2025}. Accordingly, attention-based token selection has been considered in \cite{peng2025large}. However, the method relies on heuristic importance scores and fixed top-$k$ selection rather than adapting token selection to fine-grained cross-modal alignment and the available latency budget. Consequently, existing studies still lack a lightweight and adaptive framework for selecting cross-modal tokens based on token-level cross-modal alignment under dynamic latency constraints.


        \subsection{Contributions}
            To address cross-modal token selection for latency-constrained multimodal communication, we develop a lightweight joint token selection framework based on fine-grained cross-modal alignment. A fundamental challenge is to characterize the semantic relationships among tokens from different modalities.
            A straightforward approach is to rank tokens according to their cross-attention scores \cite{peng2025large}, as commonly done using self-attention scores in single-modal token selection \cite{Liang_EViT_2022}.
            However, each cross-attention score captures only a pairwise relationship between tokens from two modalities, overlooking tokens jointly associated with multiple important cross-modal tokens. Consequently, under a limited transmission budget, pairwise ranking may discard tokens representing shared semantic concepts across modalities.

            Instead of relying on pairwise cross-attention scores, we reuse the query and key projections of the cross-attention layer to map tokens from different modalities into a common space, referred to as the shared query-key space. In this space, one modality is designated as the anchor, whose projected tokens serve as anchor queries, while the projected tokens of the remaining modalities serve as non-anchor keys. 
            This construction captures one-to-many cross-modal alignments by comparing each anchor query with multiple non-anchor keys. Inspired by germ-grain models \cite{Germ_grain}, we define a semantic grain region around each anchor query, where cross-modal alignment is quantified by angular distance and the corresponding angular threshold determines the region size. Building on this, we propose an intersection-based token selection (IBS) framework that optimizes the semantic grain region sizes under the target latency constraint and prioritizes the selection of non-anchor tokens whose keys lie within a larger number of semantic grain regions, as they are jointly aligned with more anchor queries.

            The key contributions of this work are summarized below:
            \begin{itemize}
                \item
                We propose a joint cross-modal token selection framework, coined IBS, that geometrically captures one-to-many cross-modal token relationships using angular distances in a shared query-key space and prioritizes the transmission of tokens jointly associated with a larger number of tokens from other modalities. 
                
                \item
                We formulate a latency-constrained query-specific angular threshold optimization problem to adapt the size of each anchor-centered semantic grain region to the local distribution of non-anchor keys, and solve it using a block coordinate descent (BCD) algorithm \cite{BCD} and a low-complexity greedy algorithm.

                \item
                We also develop a robust IBS (R-IBS) scheme that extends the proposed IBS framework to enable channel-aware token selection. Specifically, we model token decoding failures as outage events over Rayleigh fading channels and incorporate the resulting token-wise erasure probabilities into the geometric selection metric.

                \item 
                Through simulations on a bimodal VQA task and a trimodal audio-visual question answering (AVQA) task using the VQA v2 \cite{Goyal_VQAv2_2017} and MUSIC-AVQA v2 \cite{Liu_MUSIC_AVQA_v2_2024} datasets, we validate that the proposed IBS framework improves task accuracy over a cross-attention-based pairwise relevance
                maximization baseline by up to $27.6\%$ and $31.6\%$, respectively,
                under a latency-constrained channel. Under token-wise erasure channels, R-IBS
                achieves corresponding gains of up to $29.2\%$ and $27.1\%$ over the
                same baseline.

            \end{itemize}

            \begin{figure*}[t]
                \centering 
                {\epsfig{file=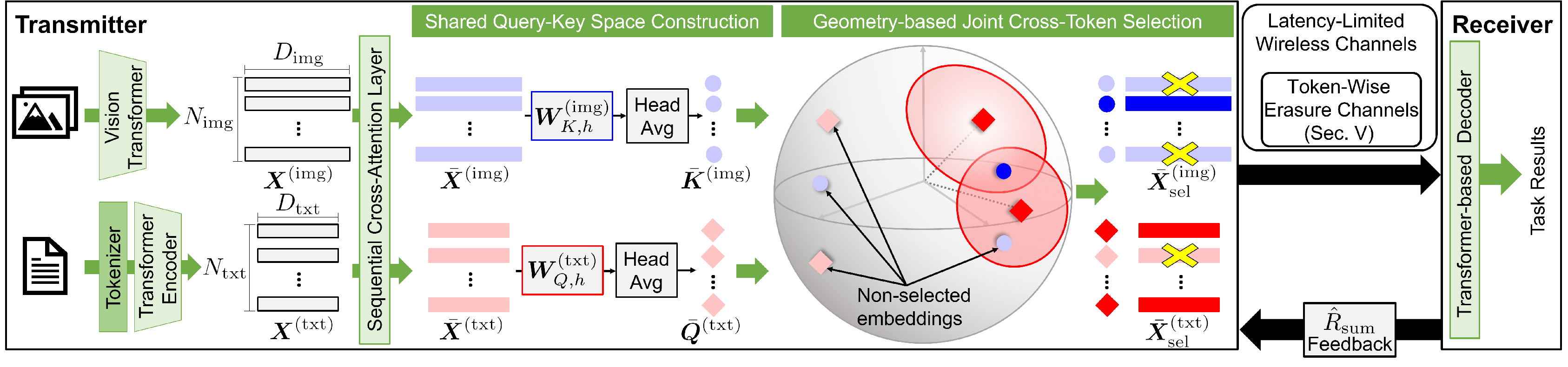, width=18cm}}
                \caption{An illustration of the proposed geometry-based joint cross-modal token selection framework for multimodal token communications.}
                \label{fig:system}
            \end{figure*}  

            This work is a substantial extension of its conference version
            \cite{Conference}, which introduced geometry-driven IBS for bimodal
            image--text token communications using a
            common angular threshold determined via bisection under a prescribed
            communication overhead. In contrast, this work generalizes IBS
            to multiple modalities through sequential cross-attention and jointly
            optimizes query-specific angular thresholds and token selection under
            an explicit latency constraint using BCD-based and low-complexity
            greedy algorithms. We further develop R-IBS for token-wise erasure
            channels by incorporating token erasure probabilities
            into the geometric selection metric.


    \section{System Model and Problem Formulation}\label{Sec:Model}

    Consider the point-to-point wireless multimodal token communication system in Fig.~\ref{fig:system}. For each streaming sample, the transmitter extracts token representations from multimodal inputs and sends them to the receiver for a downstream machine learning task. Due to a target latency constraint $T_{\rm target}$, only a fraction of generated tokens is selected for transmission.

    Let $\mathcal{M}=\{1,\dots,M\}$ denote the set of modality indices. Following Transformer-based multimodal processing \cite{Zhang_UDeepSC_2024}, the transmitter separately encodes each modality $m\in\mathcal{M}$ using a modality-specific Transformer encoder. The output of the encoder is represented as ${\boldsymbol X}^{(m)} \in \mathbb{R}^{N_m \times D_m}$,
    where $N_m$ and $D_m$ are the number of tokens and the token dimension for modality $m$, respectively. We refer to ${\boldsymbol X}^{(m)}$ as the modality-specific tokens.
    These tokens are then processed by the cross-attention mechanism, to be detailed in Sec.~\ref{Subsec:SCAL}, producing the cross-modal tokens $\{\bar{\boldsymbol X}^{(m)}\}_{m\in\mathcal{M}}$. Each $\bar{\boldsymbol X}^{(m)}\in\mathbb{R}^{N_m\times D_m}$ contains cross-modal tokens of modality $m$, which are enriched with information from other modalities.

    To satisfy the target latency constraint $T_{\rm target}$, the transmitter
    selects a fraction of the cross-modal tokens for transmission. The selected
    tokens of modality $m$ are represented by
    $\bar{\boldsymbol X}^{(m)}_{\rm sel}
    \in\mathbb{R}^{\bar{N}_m\times D_m}$, where
    $\bar{N}_m\leq N_m$ is the number of selected tokens. Each entry of
    $\bar{\boldsymbol X}^{(m)}_{\rm sel}$ is quantized using $Q_m$ bits,
    yielding $B_m=\bar{N}_mQ_mD_m$ information bits for modality $m$.
    For analytical tractability, we assume that each selected token is encoded into an independently decodable packet using a capacity-achieving channel code. The total number of transmitted bits is given by $B_{\rm sum}=\sum_{m\in\mathcal{M}}B_m$.

    We further consider a parallel transmission model comprising
    $N_{\rm sub}$ orthogonal frequency-flat subchannels. For each modality
    $m$, the coded token packets are modulated and mapped onto these
    subchannels, forming the modality-wise transmit symbol sequence 
    $\tilde{\boldsymbol s}^{(m)}=[
    ({\boldsymbol s}_1^{(m)})^\top,\ldots,
    ({\boldsymbol s}_{N_{\rm sub}}^{(m)})^\top
    ]^\top,$
    where
    ${\boldsymbol s}^{(m)}_\ell
    \in\mathbb{C}^{L^{(m)}_\ell}$
    is the symbol sequence transmitted over the $\ell$-th subchannel for
    modality $m$, and
    $L_{\rm tot}^{(m)}
    =\sum_{\ell=1}^{N_{\rm sub}}L_\ell^{(m)}$
    is the corresponding transmission block length.


    Assuming that each modality-specific transmission occurs within a
    single channel coherence interval, the received signal over the
    $\ell$-th subchannel for modality $m$ is given by
    \begin{align}
    {\boldsymbol y}_\ell^{(m)}
    =
    \sqrt{P_\ell^{(m)}}h_\ell^{(m)}
    {\boldsymbol s}_\ell^{(m)}
    +
    {\boldsymbol z}_\ell^{(m)},
    \quad
    \ell\in\{1,\ldots,N_{\rm sub}\},
    \end{align}
    where $P^{(m)}_\ell$ is the transmit power allocated to the $\ell$-th
    subchannel for modality $m$, $h^{(m)}_\ell\sim\mathcal{CN}(0,1)$ is the corresponding
    Rayleigh block-fading channel coefficient, and
    ${\boldsymbol z}^{(m)}_\ell\sim
    \mathcal{CN}({\boldsymbol 0},\sigma^2{\bf I}_{L^{(m)}_\ell})$
    is the additive white Gaussian noise vector. With
    $\mathbb{E}[\|{\boldsymbol s}_\ell^{(m)}\|^2]=L_\ell^{(m)}$, the instantaneous received SNR of the $\ell$-th subchannel for modality $m$ is given by $\gamma^{(m)}_\ell
    =
    {P^{(m)}_\ell|h^{(m)}_\ell|^2}/{\sigma^2}$.

    To enable channel-adaptive token transmission, the receiver estimates
    the channel coefficient of each subchannel using pilot symbols.
    Following \cite{Hassibi_Training_2003}, the estimated channel coefficient is represented as $\hat{h}^{(m)}_\ell=h^{(m)}_\ell + e^{(m)}_\ell$, where $e_\ell$ is the
    estimation error. Based on this, the estimated SNR of the $\ell$-th subchannel for modality $m$ is given by $\hat{\gamma}^{(m)}_\ell= {P^{(m)}_\ell|\hat{h}^{(m)}_\ell|^2}/{\sigma^2}$.
    Accordingly, the receiver computes the aggregate estimated achievable rate by averaging the modality-wise rates as
    \begin{align}
    \hat{R}_{\rm sum}
    =
    \frac{1}{M}\sum_{m\in\mathcal{M}}^{}\sum_{\ell=1}^{N_{\rm sub}}
    W_\ell\log_2(1+\hat{\gamma}^{(m)}_\ell),
    \label{eq:estimated_sum_rate}
    \end{align}
    where $W_\ell$ is the bandwidth assigned to the $\ell$-th
    subchannel, and $W$ is the total communication bandwidth, such that $\sum_{\ell=1}^{N_{\rm sub}}W_\ell=W$. 
    Prior to latency-constrained token selection, the receiver feeds back $\hat{R}_{\rm sum}$ to the transmitter through an error-free link. For analytical simplicity, $\hat{R}_{\rm sum}$ is used as a common achievable rate for all selected tokens.

    Based on the feedback $\hat{R}_{\rm sum}$, the transmitter determines
    which cross-modal tokens to transmit under the target latency constraint. The fundamental objective of latency-constrained multimodal inference is to maximize the downstream task performance at the receiver by transmitting the most informative cross-modal tokens. 
    Accordingly, the root token selection problem is formulated as
    \begin{align}
    ({\bf P1})\quad
    \max_{\{\bar{\boldsymbol X}^{(m)}_{\rm sel}\}_{m\in\mathcal{M}}}
    \quad &
    \mathcal{U}\!\left(
    \{\bar{\boldsymbol X}^{(m)}_{\rm sel}\}_{m\in\mathcal{M}}
    \right),
    \label{eq:P1_objective}
    \\
    \text{s.t.}\quad ~~~~~&
    \frac{B_{\rm sum}}{\hat{R}_{\rm sum}}
    \leq
    T_{\rm target},
    \label{eq:P1_latency_constraint}
    \\
    &
    \bar{N}_m\leq N_m,
    \quad \forall m\in\mathcal{M},
    \label{eq:P1_token_constraint}
    \end{align}
    where $\mathcal{U}(\cdot)$ denotes the downstream task utility achieved by the selected cross-modal tokens at the receiver.

    In Sec.~\ref{Sec:Channel_Aware_Metrics}, we address imperfect channel knowledge by considering a mismatch between the feedback aggregate rate and the actual achievable rate. Although the selected cross-modal tokens satisfy the target latency constraint based on $\hat{R}_{\rm sum}$, channel estimation errors and temporal channel variations may prevent some selected tokens from being reliably decoded within $T_{\rm target}$. These decoding failures are modeled as token-wise erasures, whose probabilities are incorporated into the channel-aware selection metric.

    \section{Proposed Shared Query-Key Space \\ Construction via Cross-Attention}\label{Sec:QKSpace}
    This section establishes the shared query-key space that serves as a key enabler of the proposed geometric cross-modal token selection framework. To this end, we first introduce a sequential cross-attention layer which generates modality-specific cross-modal tokens by capturing inter-modal dependencies (Sec.~\ref{Subsec:SCAL}). We then project the resulting cross-modal tokens into a unified query-key space and define a distance metric for joint cross-modal token selection (Sec.~\ref{Subsec:Shared_Space}).

    \subsection{Sequential Cross-Attention Layer}\label{Subsec:SCAL} 
    We propose a sequential cross-attention layer that applies the standard multi-head cross-attention mechanism \cite{Lu_ViLBERT_2019} once per modality. The modalities are processed in a fixed order, with each serving as the query once. At each stage, the update uses the cross-modal tokens of previously processed modalities and the original tokens of the remaining modalities, thereby progressively propagating cross-modal information. Unlike the parallel design in \cite{Lu_ViLBERT_2019}, which updates each modality independently using only the original tokens of the others, the proposed design produces progressively enriched cross-modal tokens with the same number of cross-attention operations. Its inputs are the token sequences generated by $M$ modality-specific feature extractors (e.g., a vision Transformer \cite{ViT_model}).

\begin{figure}[t]
    \centering   
    {\epsfig{file=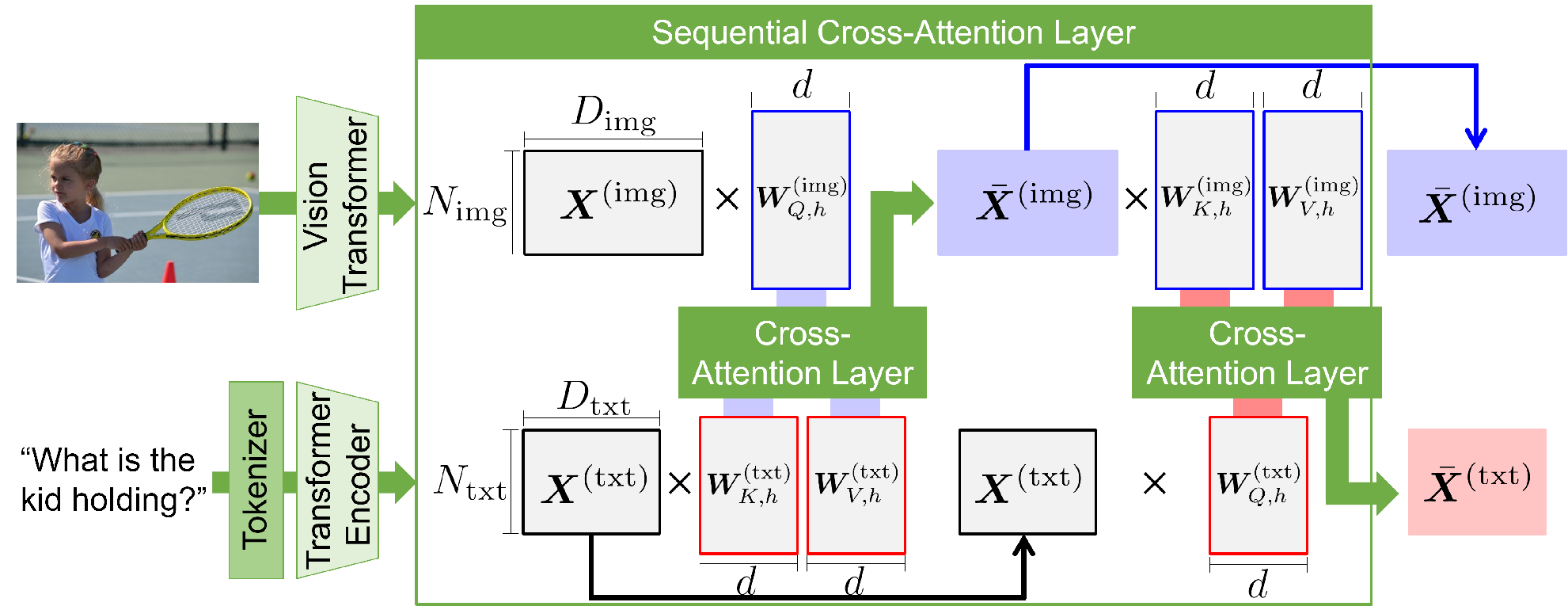, width=8.7cm}}
    \caption{
    An illustration of the proposed sequential cross-attention layer for the VQA task with image and text inputs.}
    \label{fig:system_2}
    \end{figure}

    To explain the operation of the proposed layer, we first consider the two-modality case with image and text inputs, as illustrated in Fig.~\ref{fig:system_2}. The proposed layer updates the two modalities sequentially. In the first stage, the image tokens are updated by attending to the text tokens. In the second stage, the text tokens are updated by attending to the updated image-side tokens. We begin with the image update stage, where the image tokens serve as queries and the text tokens provide the corresponding keys and values. Let $m\in\{{\rm img},{\rm txt}\}$ and $H$ be the number of attention heads. For the $h$-th head, the projected query, key, and value matrices are given by
    \begin{align}
        {\boldsymbol Q}_h^{(\rm img)}
        &= {\boldsymbol X}^{(\rm img)}{\boldsymbol W}_{Q,h}^{(\rm img)} + {\boldsymbol B}_{Q,h}^{(\rm img)} \in \mathbb{R}^{N_{\rm img} \times d}, \\
        {\boldsymbol K}_h^{(\rm txt)}
        &= {\boldsymbol X}^{(\rm txt)}{\boldsymbol W}_{K,h}^{(\rm txt)} + {\boldsymbol B}_{K,h}^{(\rm txt)} \in \mathbb{R}^{N_{\rm txt} \times d}, \\
        {\boldsymbol V}_h^{(\rm txt)}
        &= {\boldsymbol X}^{(\rm txt)}{\boldsymbol W}_{V,h}^{(\rm txt)} + {\boldsymbol B}_{V,h}^{(\rm txt)} \in \mathbb{R}^{N_{\rm txt} \times d},
    \end{align}
   where ${\boldsymbol W}_{a,h}^{(m)}$ and ${\boldsymbol B}_{a,h}^{(m)}$ are the learnable projection matrices and bias matrices, respectively, for $(m,a) \in \{({\rm img},Q), ({\rm txt},K), ({\rm txt},V)\}$, and $d$ is the feature dimension of each attention head. Using these projected matrices, the image-to-text cross-attention scores are computed as
    \begin{align}
        \!\!\!\!\!{\boldsymbol C}_h^{({\rm img}\to {\rm txt})}
        \!=\!
        \mathrm{Softmax}\!\left(\!\!
        \frac{{\boldsymbol Q}_h^{(\rm img)}({\boldsymbol K}_h^{(\rm txt)})^{\! \top}}{\sqrt{d}}\!\!
        \right)
        \!\! \in \! \mathbb{R}^{N_{\rm img} \times N_{\rm txt}}, \!\!\label{attn_score_i2t}
    \end{align}
    and the corresponding head output is obtained as
    \begin{align}
        {\boldsymbol C}_h^{(\rm img)}
        =
        {\boldsymbol C}_h^{({\rm img}\to {\rm txt})}{\boldsymbol V}_h^{(\rm txt)}
        \in \mathbb{R}^{N_{\rm img} \times d}. \label{attn_score_img}
    \end{align}
     After concatenating the outputs of all heads, the image-side tokens are updated by
    \begin{align}
        {\boldsymbol U}^{(\rm img)}
        \!&= \!\mathrm{LN}\Big(
        {\boldsymbol X}^{(\rm img)}
        + \mathrm{DP}\Big(\!
        \left[{\boldsymbol C}_1^{(\rm img)};\!\cdots\!;{\boldsymbol C}_{H}^{(\rm img)}\right]\!
        \!{\boldsymbol W}_O^{(\rm img)}
        \Big)\!
        \Big), \\
        \bar{\boldsymbol X}^{(\rm img)}
        \!&=\! \mathrm{LN}\!\left(
        {\boldsymbol U}^{(\rm img)}
        + \mathrm{DP}\!\left(
        \mathrm{FFN}\!\left({\boldsymbol U}^{(\rm img)}\!\right)\!
        \right)\!
        \right),
    \end{align}
    where ${\boldsymbol W}_O^{(\rm img)} \in \mathbb{R}^{(Hd) \times D_{\rm img}}$ is the output projection matrix. Here, $\mathrm{LN}$, $\mathrm{DP}$, and $\mathrm{FFN}$ are layer normalization, dropout, and the feed-forward network, respectively. The text update stage is then applied using the updated image-side tokens $\bar{\boldsymbol X}^{(\rm img)}$ as the source. In this stage, the text tokens serve as queries, while the updated image-side tokens provide the corresponding keys and values. For the $h$-th attention head, the projected query, key, and value matrices are given by
    \begin{align}
        {\boldsymbol Q}_h^{(\rm txt)}
        &= {\boldsymbol X}^{(\rm txt)}{\boldsymbol W}_{Q,h}^{(\rm txt)} + {\boldsymbol B}_{Q,h}^{(\rm txt)} \in \mathbb{R}^{N_{\rm txt} \times d}, \\
        {\boldsymbol K}_h^{(\rm img)}
        &= \bar{\boldsymbol X}^{(\rm img)}{\boldsymbol W}_{K,h}^{(\rm img)} + {\boldsymbol B}_{K,h}^{(\rm img)} \in \mathbb{R}^{N_{\rm img} \times d}, \\
        {\boldsymbol V}_h^{(\rm img)}
        &= \bar{\boldsymbol X}^{(\rm img)}{\boldsymbol W}_{V,h}^{(\rm img)} + {\boldsymbol B}_{V,h}^{(\rm img)} \in \mathbb{R}^{N_{\rm img} \times d},
    \end{align}
    where ${\boldsymbol W}_{a,h}^{(m)}$ and ${\boldsymbol B}_{a,h}^{(m)}$ are the learnable projection matrices and bias matrices, respectively, for $(m,a) \in \{({\rm txt},Q), ({\rm img},K), ({\rm img},V)\}$. Using these projected matrices, the text-to-image cross-attention scores are computed as
    \begin{align}
        \!\!\!\!\!{\boldsymbol C}_h^{({\rm txt}\to {\rm img})}
        \!=\!
        \mathrm{Softmax}\!\left(\!\!
        \frac{{\boldsymbol Q}_h^{(\rm txt)}({\boldsymbol K}_h^{(\rm img)})^{\!\top}}{\sqrt{d}}
        \!\!\right)
        \!\!\in \! \mathbb{R}^{N_{\rm txt} \times N_{\rm img}},\!\! \label{attn_score_t2i}
    \end{align}
    and the corresponding head output is obtained as
    \begin{align}
        {\boldsymbol C}_h^{(\rm txt)}
        =
        {\boldsymbol C}_h^{({\rm txt}\to {\rm img})}{\boldsymbol V}_h^{(\rm img)}
        \in \mathbb{R}^{N_{\rm txt} \times d}. \label{attn_score_txt}
    \end{align}
    After concatenating the outputs of all heads, the text-side tokens are updated as
    \begin{align}
        {\boldsymbol U}^{(\rm txt)}
        \!&=\! \mathrm{LN}\Big(
        {\boldsymbol X}^{(\rm txt)}
        + \mathrm{DP}\Big(\!
        \left[{\boldsymbol C}_1^{(\rm txt)};\!\cdots\!;{\boldsymbol C}_{H}^{(\rm txt)}\right]\!
        {\boldsymbol W}_O^{(\rm txt)}
        \Big)\!
        \Big), \\
        \bar{\boldsymbol X}^{(\rm txt)}
        \!&=\! \mathrm{LN}\!\left(
        {\boldsymbol U}^{(\rm txt)}
        + \mathrm{DP}\!\left(
        \mathrm{FFN}\!\left({\boldsymbol U}^{(\rm txt)}\!\right)\!
        \right)\!
        \right),
    \end{align}
    where ${\boldsymbol W}_O^{(\rm txt)} \in \mathbb{R}^{(Hd) \times D_{\rm txt}}$ is the output projection matrix. The resulting outputs $\bar{\boldsymbol X}^{(\rm txt)}$ and $\bar{\boldsymbol X}^{(\rm img)}$ are referred to as the text and image cross-modal tokens, respectively.

    For $M>2$, the modalities are sequentially updated. The case of $M=3$
    will be considered in Sec.~\ref{Sec:Simul}. 
    When modality $m$
    is updated, its tokens serve as queries, while the tokens from each of
    the other modalities $m'\in\mathcal{M}\setminus\{m\}$ serve as the
    corresponding keys and values. If modality $m'$ has already been updated
    at an earlier stage, its cross-modal tokens
    $\bar{\boldsymbol X}^{(m')}$ are used. Otherwise, its original tokens
    ${\boldsymbol X}^{(m')}$ are used. Since this sequential cross-attention
    operation is not permutation invariant, the resulting cross-modal tokens
    may depend on the update order. Accordingly, a fixed update order is
    maintained throughout both training and evaluation.


   \subsection{Shared Query-Key Space Construction}\label{Subsec:Shared_Space}

    This subsection constructs a common representation space for measuring token-level relationships across modalities, which is subsequently used for cross-modal token selection. Although the cross-modal tokens $\{\bar{\boldsymbol X}^{(m)}\}_{m\in\mathcal{M}}$ contain information
    exchanged through cross-attention, their raw embeddings remain modality-specific and are therefore not directly comparable. In particular, different modalities may employ different feature dimensions. Moreover, even when the feature dimensions are matched, high-dimensional distances can cause a few generic tokens to appear close to many others. This hubness effect reduces the reliability of raw embedding distances for cross-modal alignment \cite{Radovanovic_Hubness_2010}.
    
    To compare cross-modal tokens in a common representation space, we reuse the query and key projection matrices introduced in Sec.~\ref{Subsec:SCAL}. Specifically, the projections parameterized by ${\boldsymbol W}^{(m)}_{a,h}$ and ${\boldsymbol B}^{(m)}_{a,h}$ for $a\in\{Q,K\}$ transform modality-specific cross-modal tokens into $d$-dimensional query and key representations. In the cross-attention mechanism, the relation between two tokens is evaluated from the similarity between their query and key representations. Therefore, reusing these learned projections enables tokens from different modalities to be compared in the same space without introducing additional projection layers. This common representation space is referred to as the shared query-key space.


    Within the shared query-key space, we designate the modality with the
    smallest number of tokens as the reference modality, denoted by
    $m^\star\in\mathcal{M}$. Its projected cross-modal tokens serve as
    anchor queries, while those of the remaining modalities serve as
    non-anchor keys. This anchor-based design reduces the number of cross-modal comparisons considered in the proposed selection framework, since only the tokens of the smallest modality are used as anchors.
    To characterize the relationships between these anchor queries and
    non-anchor keys, we exploit the query and key projections learned
    through scaled dot-product attention, where their inner products
    determine token-level cross-modal interactions. This motivates the
    following working hypothesis:
    
    \vspace{1mm}
    \noindent \textbf{Hypothesis 1} (Geometric alignment hypothesis). \textit{Higher cosine similarity between an anchor query and a non-anchor key is associated with greater downstream task utility.}
    \vspace{1mm}
    
    Under the above hypothesis, cosine similarity is adopted as the geometric
    alignment metric. Empirical support for this hypothesis is provided in Sec.~\ref{Sec:Simul}.


    To construct this space, the output cross-modal tokens from the last stage of the sequential cross-attention layer are projected into anchor query and non-anchor key representations. The projected representations are then averaged across attention heads as
    \begin{align}
        \bar{\boldsymbol Q}^{(m^\star)}
        &= \sum_{h=1}^{H}
        \frac{1}{H}
        \left(
        \bar{\boldsymbol X}^{(m^\star)}{\boldsymbol W}_{Q,h}^{(m^\star)}
        + {\boldsymbol B}_{Q,h}^{(m^\star)}
        \right) \notag\\
        &= \bar{\boldsymbol X}^{(m^\star)}
        \underbrace{\frac{\sum_{h=1}^{H}{\boldsymbol W}_{Q,h}^{(m^\star)}}{H}}_{\bar{\boldsymbol W}_{Q}^{(m^\star)}}
        +
        \underbrace{\frac{\sum_{h=1}^{H}{\boldsymbol B}_{Q,h}^{(m^\star)}}{H}}_{\bar{\boldsymbol B}_{Q}^{(m^\star)}},
        \label{head_avg_q_general}
    \end{align}
    and, for each modality $m' \in \mathcal{M}\setminus\{m^\star\}$,
    \begin{align}
        \bar{\boldsymbol K}^{(m')}
        &= \sum_{h=1}^{H}
        \frac{1}{H}
        \left(
        \bar{\boldsymbol X}^{(m')}{\boldsymbol W}_{K,h}^{(m')}
        + {\boldsymbol B}_{K,h}^{(m')}
        \right) \notag\\
        &= \bar{\boldsymbol X}^{(m')}
        \underbrace{\frac{\sum_{h=1}^{H}{\boldsymbol W}_{K,h}^{(m')}}{H}}_{\bar{\boldsymbol W}_{K}^{(m')}}
        +
        \underbrace{\frac{\sum_{h=1}^{H}{\boldsymbol B}_{K,h}^{(m')}}{H}}_{\bar{\boldsymbol B}_{K}^{(m')}},
        \label{head_avg_k_general}
    \end{align}
    where $\bar{\boldsymbol Q}^{(m^\star)}\in\mathbb{R}^{N_{m^\star}\times d}$ and $\bar{\boldsymbol K}^{(m')}\in\mathbb{R}^{N_{m'}\times d}$ are the head-averaged query and key representations of the anchor modality $m^\star$ and a non-anchor modality $m'$, respectively.

    To quantify the cross-modal alignment between the anchor modality $m^\star$ and each non-anchor modality $m' \in \mathcal{M}\setminus\{m^\star\}$, we define the cosine similarity between the $i$-th anchor query and the $j$-th non-anchor key of modality $m'$ as
    \begin{align}
        S^{(m')}_{i,j}
        =
        \frac{
        \bar{\boldsymbol q}_i^{(m^\star)}
        \left(\bar{\boldsymbol k}_j^{(m')}\right)^\top
        }
        {
        \|\bar{\boldsymbol q}_i^{(m^\star)}\|_2
        \|\bar{\boldsymbol k}_j^{(m')}\|_2
        },
        \label{CS_def}
    \end{align}
    where $\bar{\boldsymbol q}_i^{(m^\star)} = [\bar{\boldsymbol Q}^{(m^\star)}]_{i,:}$ and $\bar{\boldsymbol k}_j^{(m')} = [\bar{\boldsymbol K}^{(m')}]_{j,:}$ are the $i$-th anchor query and the $j$-th non-anchor key of modality $m'$, respectively, for $i\in\{1,\ldots,N_{m^\star}\}$ and $j\in\{1,\ldots,N_{m'}\}$. To induce a geometric metric space, we define the corresponding angular distance as
    \begin{align}
        \Theta^{(m')}_{i,j}
        =
        \arccos\!\left(S^{(m')}_{i,j}\right).
        \label{ang_dist}
    \end{align}
    Based on this distance, a {\em semantic grain region} is defined for each anchor query using a query-specific angular threshold. Let $\theta_i \in [0,\pi]$ denote the angular threshold assigned to the $i$-th anchor query. Then, the corresponding grain region is defined as
    \begin{align} 
    \mathcal{G}_i(\theta_i) = \Big\{ &(m',j) ~\Big|~   \Theta^{(m')}_{i,j} \le \theta_i \Big\}. \label{grain_query_specific_theta} 
    \end{align}
    Here, $\mathcal{G}_i(\theta_i)$ contains all non-anchor keys whose angular distances from the $i$-th anchor query do not exceed its query-specific threshold $\theta_i$. 
    The overall construction of the shared query-key space and its role in the proposed framework are illustrated in Fig.~\ref{fig:system}.

    \section{Proposed Geometry-Based Joint Cross-Modal Token Selection Framework}\label{Sec:Proposed_Framework}
 
    This section presents the proposed geometry-based joint cross-modal token selection framework using semantic grain regions in the shared query-key space. The framework exploits the geometric overlap between anchor queries and non-anchor keys, prioritizing non-anchor tokens whose key representations are covered by multiple anchor-centric semantic grain regions as shared semantic evidence for transmission.
    
    To account for the non-uniform local geometry of the shared query-key space, we optimize the query-specific angular thresholds under the target latency constraint. Each threshold controls the size of the grain region associated with its anchor query according to the local distribution of non-anchor keys. A low-complexity greedy algorithm is further developed to reduce the computational burden of the exact optimization-based solution. For ease of exposition, we assume error-free transmission in this section, and extend the framework to imperfect channels in Sec.~\ref{Sec:Channel_Aware_Metrics}.

    \subsection{Intersection-Based Cross-Modal Token Selection}\label{Subsec:IBS}


    Building on the geometric alignment hypothesis and semantic grain
    construction in Sec.~\ref{Subsec:Shared_Space}, we adopt the following
    working hypothesis:

    \vspace{1mm}
    \noindent \textbf{Hypothesis 2} (Multi-query consensus hypothesis). \textit{A non-anchor key jointly aligned with multiple anchor queries provides more reliable evidence of shared cross-modal relevance than one aligned with only a single anchor query.}
    \vspace{1mm}
    
    Based on this hypothesis, we propose IBS, which prioritizes non-anchor
    keys lying in multiple anchor-centric semantic grain regions.
    Empirical support for this hypothesis is provided in
    Sec.~\ref{Sec:Simul} through comparison with cross-attention-based
    pairwise relevance maximization.

    To formalize this rule, let $\boldsymbol{\theta}=\{\theta_i\}_{i=1}^{N_{m^\star}}$. For any anchor index set $\mathcal{A}\subseteq\{1,\dots,N_{m^\star}\}$, we define the intersection of the corresponding semantic grain regions as 
    \begin{align} 
    \mathcal{G}_{\mathcal{A}}(\boldsymbol{\theta}) = \bigcap_{i\in\mathcal{A}} \mathcal{G}_i(\theta_i). 
    \label{eq:grain_intersection} 
    \end{align} 
    A non-anchor key is regarded as satisfying the IBS condition if it belongs to the intersection of at least $k$ semantic grain regions, where $k\ge 2$ is a prescribed overlap threshold. Thus, the selected non-anchor token set is defined as 
    \begin{align} 
    \mathcal{S}_{\rm IBS}^{(\setminus m^\star)}(\boldsymbol{\theta};k) = \bigcup_{\substack{ \mathcal{A}\subseteq\{1,\dots,N_{m^\star}\!\},~|\mathcal{A}|=k }} \mathcal{G}_{\mathcal{A}}(\boldsymbol{\theta}). 
    \label{eq:IBS_nonanchor_set} 
    \end{align} 
    The corresponding selected anchor token set is obtained by collecting anchor queries whose semantic grain regions contain at least one selected non-anchor key, i.e., 
    \begin{align} 
    \mathcal{S}_{\rm IBS}^{(m^\star)}(\boldsymbol{\theta};k)\! = \!\Big\{ i \in \{1,\dots,N_{m^\star}\} \Big| \mathcal{G}_i(\theta_i) \cap \mathcal{S}_{\rm IBS}^{(\setminus m^\star)}(\boldsymbol{\theta};k) \neq \emptyset \Big\}. 
    \label{eq:IBS_anchor_set} 
    \end{align}

    \subsection{Joint Cross-Modal Token Selection Problem with Angular Threshold Optimization}\label{Subsec:Adaptive_Theta}
    The problem $({\bf P1})$ aims to maximize the generic downstream task
    utility $\mathcal{U}(\cdot)$ under the target latency constraint.
    However, $\mathcal{U}(\cdot)$ depends on the receiver-side inference
    model and generally does not admit an explicit form that can be
    optimized before transmission. We therefore construct a tractable
    surrogate using the geometric alignment and multi-query consensus
    hypotheses introduced in Sec.~\ref{Subsec:Shared_Space} and
    Sec.~\ref{Subsec:IBS}, respectively.
    Specifically, we approximate $\mathcal{U}(\cdot)$ by the sum of the
    cosine similarities in \eqref{CS_def} over selected anchor--non-anchor
    pairs satisfying the semantic grain condition. We further impose the
    multi-query consensus condition by requiring each selected non-anchor
    token to be aligned with at least $k$ selected anchor queries, yielding
    a geometry-based surrogate of $({\bf P1})$.
        
    To formulate the resulting problem, we introduce binary selection
    variables for the anchor and non-anchor tokens, along with continuous
    query-specific angular threshold variables. Let $x_i\in\{0,1\}$
    indicate whether the $i$-th anchor token is selected, and let
    $y_j^{(m')}\in\{0,1\}$ indicate whether the $j$-th non-anchor token of
    modality $m'\in\mathcal{M}\setminus\{m^\star\}$ is selected. Each
    anchor query is assigned an angular threshold
    $\theta_i\in[0,\pi]$, which determines the size of its semantic grain
    region according to the local distribution of non-anchor keys.
    The resulting geometry-based surrogate problem is formulated as
    \begin{subequations} 
    \begin{align} 
    \!\!\!\!\!\!\!({\bf P2})
    &\max_{\{\theta_i,x_i\},\,\{y_j^{(m')}\}}
    \;
    \sum_{m' \neq m^\star}
    \sum_{j=1}^{N_{m'}}
    \sum_{i=1}^{N_{m^\star}}
    S^{(m')}_{i,j}
    A^{(m')}_{i,j}
    x_i y^{(m')}_j\!\!, \!\!\!\!
    \label{Eq:Theta_Obj} \\
    \text{s.t.}~&\quad
    \sum_{i=1}^{N_{m^\star}}
    A^{(m')}_{i,j} x_i
    \ge k\,y^{(m')}_j,
    \quad \forall m',j,
    \label{Eq:Theta_C1} \\
    &
    B_{\rm sum}(\{x_i\}, \{y^{(m')}_j\}) \leq T_{\rm target}\hat{R}_{\rm sum},
    \label{Eq:Theta_C2} \\
    &
    x_i\in\{0,1\},\quad \forall i,
    \label{Eq:Theta_C3} \\
    &
    y^{(m')}_j\in\{0,1\},
    \quad  \forall m',j,
    \label{Eq:Theta_C4} \\
    &
    \theta_i\in[0,\pi],
    \quad \forall i,
    \label{Eq:Theta_C5}
    \end{align} 
    \end{subequations}
    where $B_{\rm sum}(\{x_i\}, \{y^{(m')}_j\}) = Q_{m^\star}D_{m^\star}
    \sum_{i=1}^{N_{m^\star}}x_i
    + \sum_{m' \neq m^\star} 
    Q_{m'}D_{m'}
    \sum_{j=1}^{N_{m'}}y^{(m')}_j$, $A^{(m')}_{i,j} = \mathbb{I}\big(\Theta^{(m')}_{i,j} \leq \theta_i\big)$ is the geometric inclusion indicator, and $k \ge 2$ is a prescribed overlap threshold. 
    The objective in \eqref{Eq:Theta_Obj} serves as a tractable surrogate for the utility $\mathcal{U}(\cdot)$ in $(\mathbf{P}1)$. Constraint \eqref{Eq:Theta_C1} enforces the geometric cross-modal token selection rule by requiring each transmitted non-anchor token (i.e., $y^{(m')}_j=1$) to lie within the semantic grain regions of at least $k$ selected anchor queries. Constraint \eqref{Eq:Theta_C2} ensures that the transmission latency does not exceed $T_{\rm target}$.

    \subsection{Block Coordinate Descent Algorithm for Joint Cross-Modal Token Selection}\label{Subsec:Adaptive_Theta_BCD}

    In problem $({\bf P2})$, the binary selection variables $\{x_i\}_{i=1}^{N_{m^\star}}$ and $\{y_j^{(m')}\}_{j=1}^{N_{m'}},\, m' \in \mathcal{M}\setminus\{m^\star\}$ are coupled with the query-specific continuous angular thresholds $\{\theta_i\}_{i=1}^{N_{m^\star}}$ through the non-differentiable indicator function $\mathbb{I}(\cdot)$. Together with the generalized IBS constraint in \eqref{Eq:Theta_C1}, this makes $({\bf P2})$ a non-convex MINLP problem that is generally NP-hard. To address this difficulty, we adopt the BCD algorithm \cite{BCD}, which updates one block of variables at a time while keeping the others fixed. This decomposition yields tractable subproblems and a monotonically non-decreasing objective sequence. Specifically, at iteration $t \in \{1,2,\dots,T_{\rm max}\}$, the continuous variables $\{\theta_i^{(t+1)}\}$ and the discrete variables $\{x_i^{(t+1)},y_j^{(m',t+1)}\}$ are alternately updated from the optimized variables at iteration $t$ as follows.
    \begin{itemize}
    \item {\bf (X,Y)-update:}
    With the query-specific angular thresholds fixed at $\{\theta_i^{(t)}\}$, the geometric inclusion indicator becomes a deterministic binary constant given by $A_{i,j}^{(m',t)} = \mathbb{I}\big(\Theta^{(m')}_{i,j} \le \theta_i^{(t)}\big)$.
    Accordingly, the effective similarity contribution of the $i$-th anchor query to the $j$-th non-anchor key is represented by the fixed weight $\omega_{i,j}^{(m',t)} = S^{(m')}_{i,j} A_{i,j}^{(m',t)}$.
    Under this setting, the objective function of $({\bf P2})$ becomes bilinear in the binary variables $\{x_i\}_{i=1}^{N_{m^\star}}$ and $\{y_j^{(m')}\}_{j=1}^{N_{m'}}$ . To obtain a tractable reformulation, we introduce an auxiliary continuous variable $v_j^{(m')} = y_j^{(m')} \bar{\omega}_j^{(m',t)} \in \mathbb{R}$, where $\bar{\omega}_j^{(m',t)} = \sum_{i=1}^{N_{m^\star}} \omega_{i,j}^{(m',t)}x_i$, and linearize the resulting product using the Big-$M$ method. The corresponding tight bounds are given by 
    $\underline{\Omega}_j^{(m',t)}
    \triangleq
    \sum_{i:\,\omega_{i,j}^{(m',t)}<0}
    \omega_{i,j}^{(m',t)}, ~
    \overline{\Omega}_j^{(m',t)}
    \triangleq
    \sum_{i:\,\omega_{i,j}^{(m',t)}>0}
    \omega_{i,j}^{(m',t)}$. 
    For notational simplicity, let $\tilde y_j^{(m')}=1-y_j^{(m')}$.
    Using these bounds, the resulting mixed-integer linear programming (MILP) problem is formulated as
    \begin{subequations}
    \begin{align}
    ({\bf P{2.1}})\;
    &\max_{\{x_i\},\,\{y_j^{(m')},v_j^{(m')}\}}
    \quad
    \sum_{m'\neq m^\star}
    \sum_{j=1}^{N_{m'}}
    v_j^{(m')},
    \label{Eq:XY_Obj}\\
    \text{s.t.}\quad&
    \underline{V}_j^{(m',t)}
    \le
    v_j^{(m')}
    \le
    \overline{V}_j^{(m',t)},
    \quad \forall m',j,
    \label{Eq:XY_C2}\\
    &
    \eqref{Eq:Theta_C1}\text{--}\eqref{Eq:Theta_C4},
    \notag
    \end{align}
    \end{subequations}
    where
    \begin{align}
    \underline{V}_j^{(m',t)}
    &=
    \max\!\left\{
    \bar{\omega}_j^{(m',t)}
    -
    \overline{\Omega}_j^{(m',t)}
    \tilde y_j^{(m')},
    \,
    \underline{\Omega}_j^{(m',t)}
    y_j^{(m')}
    \right\}, \notag\\
    \overline{V}_j^{(m',t)}
    &=
    \min\!\left\{
    \bar{\omega}_j^{(m',t)}
    -
    \underline{\Omega}_j^{(m',t)}
    \tilde y_j^{(m')},
    \,
    \overline{\Omega}_j^{(m',t)}
    y_j^{(m')}
    \right\}.
    \label{Eq:XY_Bounds}
    \end{align}

    Since $({\bf P{2.1}})$ is a MILP with linear objective and constraint functions over a bounded feasible region, it admits a globally optimal solution. Therefore, standard branch-and-cut methods \cite{Padberg_BranchAndCut_1991} can be used to obtain the global optimum of $({\bf P{2.1}})$. Accordingly, the optimal binary variables are updated as $\{x_i^{(t+1)}, y_j^{(m', t+1)}\}$.
    
    \item {$\boldsymbol\theta$\bf-update:}
    With the binary cross-modal token indicators fixed at their updated values $\{x_i^{(t+1)}\}$ and $\{y_j^{(m',t+1)}\}$, the selected anchor and non-anchor token sets are defined as $\mathcal{S}_{\rm A}^{(t+1)} \triangleq \{i \mid x_i^{(t+1)}=1\}$ and $\mathcal{S}_{\rm N}^{(t+1)} \triangleq \{(m',j) \mid y_j^{(m',t+1)}=1\}$.
    Although each query-specific threshold $\theta_i$ is defined over the continuous interval $[0,\pi]$, the geometric inclusion indicator changes only at the discrete angular distances ${\Theta_{i,j}^{(m')}}$ between the anchor queries and non-anchor keys. Therefore, the continuous optimization over ${\theta_i}$ can be equivalently reduced to a finite combinatorial problem over these observed distances. In this reduced problem, we directly optimize the binary inclusion indicators ${A_{i,j}^{(m')}}$ in \eqref{Eq:Theta_Obj}, which represent whether each non-anchor key is included in the semantic grain region of each anchor query. The threshold update problem is then formulated as
    \begin{subequations}
    \begin{align}
        \!\!\!\!({\bf P{2.2}})&\max_{\{A_{i,j}^{(m')}\}}
        \sum_{(m',j) \in \mathcal{S}_{\rm N}^{(t+1)}} \sum_{i \in \mathcal{S}_{\rm A}^{(t+1)}}
        S_{i,j}^{(m')} A_{i,j}^{(m')},
        \label{Eq:A_Obj} \\
        \text{s.t.}&
         \sum_{i \in \mathcal{S}_{\rm A}^{(t+1)}} A_{i,j}^{(m')} \ge k,
        ~
        \forall (m',j) \in \mathcal{S}_{\rm N}^{(t+1)},
        \label{Eq:A_C1} \\
        & A_{i,j_1}^{(m')} \ge A_{i,j_2}^{(m')}, \quad \text{if}~ \Theta_{i,j_1}^{(m')} < \Theta_{i,j_2}^{(m')}, \notag\\ 
        & ~~\forall i \in \mathcal{S}_{\rm A}^{(t+1)},~ \forall (m',j_1),(m',j_2) \in \mathcal{S}_{\rm N}^{(t+1)}\!\!. \label{Eq:A_C2} \\
        & A_{i,j}^{(m')} \in \{0,1\},
        ~
        \forall i {\in} \mathcal{S}_{\rm A}^{(t+1)}\!\!\!,~
        \forall (m',j) \in \mathcal{S}_{\rm N}^{(t+1)}\!\!\!.
        \label{Eq:A_C3}
    \end{align}
    \end{subequations}
    Constraint \eqref{Eq:A_C2} enforces radial monotonicity. Specifically, if an anchor query includes a farther non-anchor key within its semantic grain region, then all closer non-anchor keys must also be included. Since both the objective and the constraints of $({\bf P{2.2}})$ are linear in the binary decision variables, this subproblem is a standard integer linear programming (ILP) problem. Accordingly, an optimal binary assignment, denoted by $\{A_{i,j}^{(m',t+1)}\}$, can be efficiently obtained using standard ILP solvers \cite{Nemhauser_ILP_1988}. Once the optimal assignment matrix is obtained, the corresponding query-specific angular thresholds are recovered by setting each threshold to the largest angular distance among the enclosed non-anchor keys, i.e.,
    \begin{align} 
    \theta_i^{(t+1)} = 
    \max_{\substack{ (m',j) \in \mathcal{S}_{\rm N}^{(t+1)},\\ A_{i,j}^{(m',t+1)}=1 }} \Theta_{i,j}^{(m')}, \quad \forall i \in \mathcal{S}_{\rm A}^{(t+1)}. 
    \label{Eq:Theta_Recovery} 
    \end{align}

    \end{itemize}

    Although the original problem $({\bf P2})$ is a non-convex MINLP, the proposed BCD algorithm exhibits reliable convergence behavior. In particular, each iteration consists of two block updates, namely the \textbf{(X,Y)-update} and the \textbf{$\boldsymbol{\theta}$-update}, each of which solves the corresponding subproblem to global optimality. Since each block is optimized over its feasible set while the remaining variables are fixed, the objective value cannot decrease after either update. Moreover, the objective in \eqref{Eq:Theta_Obj} is upper-bounded because it consists of a finite sum of cosine similarity terms under the target latency $T_{\rm target}$. Therefore, the objective sequence generated by the proposed BCD iterations is monotonically non-decreasing and upper-bounded, which guarantees convergence of the objective value. Consequently, the proposed BCD algorithm reliably converges to a near-optimal solution.

    \subsection{Low-Complexity Greedy Algorithm}\label{Subsec:Adaptive_Theta_LC}
    While the proposed BCD algorithm provides near-optimal solutions, solving the resulting MILP and ILP subproblems at each iteration incurs a substantial computational burden. Since standard solvers rely on branch-and-bound or branch-and-cut procedures, the worst-case complexity grows exponentially with the number of binary decision variables. Accordingly, the worst-case computational complexity of the exact BCD iterations can be expressed as $\mathcal{O}\!\left(T_{\rm max}\,2^{\,N_{m^\star}N_{\setminus m^\star}}\right)$,
    where $N_{\setminus m^\star} = \sum_{m' \in \mathcal{M}\setminus\{m^\star\}} N_{m'}$.
    This exponential growth becomes prohibitive in large-scale token selection scenarios. To address this limitation, we next develop a low-complexity algorithm based on a greedy approach.

    The fundamental intuition behind this approach stems from the observation in $({\bf P{2.2}})$ that the geometric inclusion indicator $A^{(m')}_{i,j}$ changes only at the observed discrete distance values. Consequently, rather than searching within the continuous angular space $[0, \pi]$, the optimization of $\theta_i$ can be naturally transformed to an integer rank selection problem. In particular, for the $i$-th anchor query, expanding its continuous semantic radius $\theta_i$ is equivalent to sequentially enclosing its nearest non-anchor keys in descending order of their cosine similarities $S^{(m')}_{i,j}$.

    The algorithm operates iteratively by maintaining a query-specific integer rank pointer $p_i$ for the $i$-th anchor query, which represents the number of nearest non-anchor keys currently enclosed within its semantic grain region. At each iteration, the algorithm evaluates the marginal efficiency of expanding the region of the $i$-th anchor query to encompass its $(p_i+1)$-th nearest non-anchor key, denoted by $(m'_i,j_i)$. To quantify the efficacy of enclosing this candidate key, we define the associated marginal cost and marginal gain. The marginal cost, which quantifies the additional cross-modal token budget incurred by expanding the semantic grain region of the $i$-th anchor query to include the $j_i$-th non-anchor key of modality $m'_i$, is given by
    \begin{align}
        \Delta {\rm Cost}_{i,j_i}^{(m'_i)}
        =
        \mathbb{I}(x_i = 0)
        +
        \mathbb{I}\left(o_{j_i}^{(m'_i)} = \tilde{k} \right),
        \label{Eq:Greedy_Cost}
    \end{align}
    where $\tilde{k} =k-1$, and $o_{j_i}^{(m'_i)}$ is the current overlap counter of the $j_i$-th non-anchor key of modality $m'_i$, i.e., the number of anchor queries whose semantic grain regions currently enclose this key. The first term accounts for the initial activation of a previously unselected anchor query. The second term assigns a unit cost precisely when $o_{j_i}^{(m'_i)}=k-1$, which corresponds to the critical transition where the proposed expansion makes the candidate non-anchor key satisfy the IBS overlap requirement and become formally selected.

    Next, the marginal gain associated with the proposed expansion is defined as
    \begin{align}
        \!\!\!\Delta {\rm Gain}_{i,j_i}^{(m'_i)}
        \!=
        \!\mathbb{I}\!\left(o_{j_i}^{(m'_i)} \ge \tilde{k}\right)
        \!S^{(m'_i)}_{i,j_i} \!+\!
        \mathbb{I}\!\left(o_{j_i}^{(m'_i)} = \tilde{k}\right)
        \!U_{j_i}^{(m'_i)}, \label{Eq:Greedy_Gain}
    \end{align}
    where $U_{j_i}^{(m'_i)}$ is the accumulated pending similarity score contributed by the $o_{j_i}^{(m'_i)}$ anchor queries whose semantic grain regions currently enclose the $j_i$-th non-anchor key of modality $m'_i$. This expression captures the three possible geometric states of the candidate non-anchor key under IBS. First, if $o_{j_i}^{(m'_i)}=k-2$, the proposed enclosure only creates a pending connection and therefore yields no immediate objective gain. Second, if $o_{j_i}^{(m'_i)}=k-1$, the proposed expansion causes the candidate non-anchor key to satisfy the IBS overlap condition and become newly selected. In this case, the resulting gain consists of both the current similarity $S^{(m'_i)}_{i,j_i}$ and the previously accumulated pending score $U_{j_i}^{(m'_i)}$. Finally, if $o_{j_i}^{(m'_i)}\ge k$, the candidate non-anchor key is already active, and the new enclosure contributes only its individual similarity $S^{(m'_i)}_{i,j_i}$ to the objective.
    Based on these two marginal quantities, the algorithm defines the cost-efficiency ratio for the candidate expansion between the $i$-th anchor query and the $j_i$-th non-anchor key of modality $m'_i$ as
    \begin{align}
        \rho_{i,j_i}^{(m'_i)}
        =
        \frac{
        \Delta {\rm Gain}_{i,j_i}^{(m'_i)}
        }{
        \Delta {\rm Cost}_{i,j_i}^{(m'_i)} + \epsilon
        },
        \label{Eq:Greedy_Ratio}
    \end{align}
    where $\epsilon$ is a negligibly small positive constant introduced to prevent division by zero.

    At each iteration, the algorithm computes $\rho_{i,j_i}^{(m'_i)}$ for all anchor queries, where $(m'_i,j_i)$ denotes the uniquely designated candidate corresponding to the $(p_i+1)$-th nearest non-anchor key of the $i$-th anchor query. It then greedily selects the anchor query that maximizes this ratio, i.e.,
    \begin{align}
        i^* = \arg\max_i \rho_{i,j_i}^{(m'_i)}.
    \end{align}
    This selection simultaneously determines the associated candidate non-anchor key $(m'_{i^*},j_{i^*})$ for expansion. After this joint selection, the corresponding binary cross-modal token indicators $(x_{i^*}, y_{j_{i^*}}^{(m'_{i^*})})$, the overlap counter $o_{j_{i^*}}^{(m'_{i^*})}$, and the accumulated pending score $U_{j_{i^*}}^{(m'_{i^*})}$ are updated to reflect the newly induced geometric configuration. The rank pointer $p_{i^*}$ is then incremented by one. 
    This procedure is repeated until no additional cross-modal token can be selected without violating the target latency constraint.

    To assess the computational efficiency of the proposed discrete greedy algorithm, we first examine the initialization stage. Computing cosine similarities over the projected dimension $d$ and sorting the $N_{\setminus m^\star}$ candidate non-anchor keys for all $N_{m^\star}$ anchor queries require $\mathcal{O}\!\left(N_{m^\star} N_{\setminus m^\star}\left(\log N_{\setminus m^\star} + d\right)\right)$
    operations. During the greedy stage, the algorithm evaluates the cost-efficiency ratios for all $N_{m^\star}$ anchor queries at each iteration. Although the algorithm typically terminates once the target latency $T_{\rm target}$ is reached, the worst-case number of iterations is upper-bounded by the total number of possible pointer advances, namely $N_{m^\star}N_{\setminus m^\star}$. Therefore, the overall worst-case time complexity is given by $\mathcal{O}\!\left(N_{m^\star} N_{\setminus m^\star}
    \left(\log N_{\setminus m^\star}+ d+ N_{m^\star}\right)\right)$.
    Compared with the exponential complexity of the exact BCD-based optimization, this polynomial-time complexity demonstrates the practical suitability of the proposed method for large-scale cross-modal token selection and resource-constrained deployment.

    \section{Robust Cross-Modal Token Selection under Communication Errors}\label{Sec:Channel_Aware_Metrics}
    The cross-modal token selection framework presented in Sec.~\ref{Sec:Proposed_Framework} assumes error-free transmission. In practical wireless systems, however, physical-layer impairments such as deep fading and decoding failures may cause cross-modal token losses, thereby degrading the cross-modal alignment between anchor queries and non-anchor keys. To address this issue, we extend IBS to account for transmission errors, yielding the R-IBS framework. Inspired by packet-level erasure channels \cite{Tian_Erasurechannel_2025, Lee_Erasurechannel_2025}, we introduce a token-wise erasure model, where each transmitted cross-modal token is either perfectly recovered or completely lost. Based on this model, we derive an expected cosine-similarity metric and develop channel-aware selection rules.

    Under the modality-wise transmission model described in
    Sec.~\ref{Sec:Model}, the actual aggregate achievable rate during
    the transmission of the token packets of modality $m$ is given by
    \begin{align}
    R_{\rm sum}^{(m)}
    =
    \sum_{\ell=1}^{N_{\rm sub}}
    W_\ell
    \log_2\left(1+\gamma_{\ell}^{(m)}\right).
    \label{eq:actual_token_rate}
    \end{align}
    Due to channel estimation errors, feedback delay, or
    other physical-layer impairments, $R_{\rm sum}^{(m)}$ may fall below
    the feedback aggregate rate $\hat{R}_{\rm sum}$. In this case, the
    corresponding token packets cannot be reliably decoded within their
    allocated transmission intervals and are therefore regarded as erased.
    The resulting token erasure probability for modality $m$ is
    characterized by the outage probability \cite{Outage_prob}, i.e.,
    \begin{align}
    p_e^{(m)}
    &=
    \Pr\left(
    R_{\rm sum}^{(m)}
    <
    \hat{R}_{\rm sum}
    \right).
    \label{eq:token_outage_probability}
    \end{align}

    For each token index $i \in \{1,\dots,N_m\}$, we define the corresponding binary reception indicator as
    \begin{align}
        \epsilon_i^{(m)} \sim \mathrm{Bernoulli}(1-p_e^{(m)}),
    \end{align}
    where $\{\epsilon_i^{(m)}\}_{\forall m, \forall i}$ are assumed to be independently distributed. Based on these token-wise indicators, we define the diagonal erasure matrix of modality $m$ as
    \begin{align}
    {\boldsymbol E}^{(m)}
    =
    \mathrm{diag}\!\left(
    \epsilon_1^{(m)},\epsilon_2^{(m)},\dots,\epsilon_{N_m}^{(m)}
    \right)
    \in \mathbb{R}^{N_m \times N_m}.
    \end{align}
    Since an erased cross-modal token is treated as a null vector at the receiver, the perturbed cross-modal tokens of modality $m$ are given by
    \begin{align}
        \tilde{\boldsymbol X}^{(m)}
        =
        {\boldsymbol E}^{(m)} \bar{\boldsymbol X}^{(m)}.
        \label{Eq:Perturbed_X_general}
    \end{align}

    To characterize the effect of cross-modal token erasures on the cross-modal geometry, we model the perturbed anchor query and non-anchor key matrices as random matrices. Based on the head-averaged projection in \eqref{head_avg_q_general}, the perturbed query matrix associated with the anchor modality $m^\star$ is written as
    \begin{align}
    \tilde{\boldsymbol Q}^{(m^\star)}
    &= \sum_{h=1}^{H}
    \frac{1}{H}
    \left(
    \tilde{\boldsymbol X}^{(m^\star)}{\boldsymbol W}_{Q,h}^{(m^\star)}
    + {\boldsymbol B}_{Q,h}^{(m^\star)}
    \right) \notag\\
    &= {\boldsymbol E}^{(m^\star)}
    \bar{\boldsymbol X}^{(m^\star)}\bar{\boldsymbol W}_{Q}^{(m^\star)}
    + \bar{\boldsymbol B}_{Q}^{(m^\star)} \notag\\
    &= {\boldsymbol E}^{(m^\star)}
    \left(
    \bar{\boldsymbol Q}^{(m^\star)} - \bar{\boldsymbol B}_{Q}^{(m^\star)}
    \right)
    + \bar{\boldsymbol B}_{Q}^{(m^\star)} \notag\\
    &= {\boldsymbol E}^{(m^\star)}\bar{\boldsymbol Q}^{(m^\star)}
    + \left({\bf I}_{N_{m^\star}}-{\boldsymbol E}^{(m^\star)}\right)
    \bar{\boldsymbol B}_{Q}^{(m^\star)},
    \label{Eq:Perturbed_Q_general}
    \end{align}
    where ${\bf I}_{N_{m^\star}}$ is the $N_{m^\star}\times N_{m^\star}$ identity matrix.
    
    Similarly, for each non-anchor modality $m' \in \mathcal{M}\setminus\{m^\star\}$, the perturbed key matrix is given by
    \begin{align}
        \tilde{\boldsymbol K}^{(m')}
        &= {\boldsymbol E}^{(m')}\bar{\boldsymbol K}^{(m')}
        + \left({\bf I}_{N_{m'}}-{\boldsymbol E}^{(m')}\right)
        \bar{\boldsymbol B}_{K}^{(m')}.
        \label{Eq:Perturbed_K_general}
    \end{align}
    
    Let $\tilde{\boldsymbol q}_i^{(m^\star)} = [\tilde{\boldsymbol Q}^{(m^\star)}]_{i,:}$ and $\tilde{\boldsymbol k}_j^{(m')} = [\tilde{\boldsymbol K}^{(m')}]_{j,:}$ denote the corresponding row vectors. In addition, let $\bar{\boldsymbol b}_{Q}^{(m^\star)} = [\bar{\boldsymbol B}_{Q}^{(m^\star)}]_{i,:}$ and $\bar{\boldsymbol b}_{K}^{(m')} = [\bar{\boldsymbol B}_{K}^{(m')}]_{j,:}$ denote the row vectors of the head-averaged bias matrices, which are identical across all rows. Then, from \eqref{Eq:Perturbed_Q_general} and \eqref{Eq:Perturbed_K_general}, we obtain
    \begin{align}
        \tilde{\boldsymbol q}_i^{(m^\star)}
        &=
        \epsilon_i^{(m^\star)} \bar{\boldsymbol q}_i^{(m^\star)}
        +
        \left(1-\epsilon_i^{(m^\star)}\right)\bar{\boldsymbol b}_{Q}^{(m^\star)},
        \label{Eq:Perturbed_q_row_general} \\
        \tilde{\boldsymbol k}_j^{(m')}
        &=
        \epsilon_j^{(m')} \bar{\boldsymbol k}_j^{(m')}
        +
        \left(1-\epsilon_j^{(m')}\right)\bar{\boldsymbol b}_{K}^{(m')}.
        \label{Eq:Perturbed_k_row_general}
    \end{align}
    
    As established in Sec.~\ref{Sec:Proposed_Framework}, the original cross-modal token selection framework defines the angular distance metric from the deterministic cosine similarity $S_{i,j}^{(m')}$ in \eqref{CS_def}. Under the stochastic cross-modal token erasure model, however, the cosine similarity becomes random due to the possible erasures of both the anchor query and the non-anchor key. To obtain a robust and computable distance metric under channel uncertainty, we consider its statistical expectation. In particular, for a non-anchor key associated with modality $m' \in \mathcal{M}\setminus\{m^\star\}$, let $\tilde{S}_{i,j}^{(m')}$ denote the resulting random cosine similarity. 
    Since the token-wise erasures of the $i$-th anchor query and the $j$-th non-anchor key are independently governed by $\epsilon_i^{(m^\star)}$ and $\epsilon_j^{(m')}$, respectively, its expectation is obtained by averaging over the four possible reception states as
    \begin{align} \mathbb{E}\!\left[\tilde{S}_{i,j}^{(m')}\right] &= (1-p_e^{(m^\star)})(1-p_e^{(m')}) \psi\!\left( \bar{\boldsymbol q}_i^{(m^\star)}, \bar{\boldsymbol k}_j^{(m')} \right) \notag\\ 
    &\quad+ (1-p_e^{(m^\star)}) p_e^{(m')} \psi\!\left( \bar{\boldsymbol q}_i^{(m^\star)}, \bar{\boldsymbol b}_{K}^{(m')} \right) \notag\\ 
    &\quad+ p_e^{(m^\star)} (1-p_e^{(m')}) \psi\!\left( \bar{\boldsymbol b}_{Q}^{(m^\star)}, \bar{\boldsymbol k}_j^{(m')} \right) \notag\\ 
    &\quad+ p_e^{(m^\star)} p_e^{(m')} \psi
    \left( \bar{\boldsymbol b}_{Q}^{(m^\star)}, \bar{\boldsymbol b}_{K}^{(m')} \right), \label{Eq:Expected_CS_general} 
    \end{align} 
    where $\psi(\boldsymbol a,\boldsymbol b) = \frac{\boldsymbol a\boldsymbol b^\top} {\|\boldsymbol a\|_2\|\boldsymbol b\|_2}$ is the cosine similarity between two vectors.
    Therefore, the proposed geometric cross-modal token selection framework in Sec.~\ref{Sec:Proposed_Framework} can be readily extended to token-wise erasure channels by replacing the original angular distance with its channel-aware counterpart. Specifically, for each non-anchor modality $m' \in \mathcal{M}\setminus\{m^\star\}$, we define the channel-aware distance matrix $\tilde{\boldsymbol \Theta}^{(m')} \in \mathbb{R}^{N_{m^\star}\times N_{m'}}$,
    whose $(i,j)$-th entry is given by
    \begin{align}
        \tilde{\Theta}_{i,j}^{(m')}
        \triangleq
        \arccos\!\left(
        \mathbb{E}\!\left[\tilde{S}_{i,j}^{(m')}\right]
        \right).
    \end{align}
    Accordingly, the semantic grain region in \eqref{grain_query_specific_theta} is extended to its channel-aware counterpart as
    \begin{align} 
    \tilde{\mathcal G}_i(\theta_i)
    = \Big\{ &(m',j) ~\Big|~  \tilde{\Theta}^{(m')}_{i,j} \le \theta_i \Big\}.
    \end{align}
    The adaptive angular threshold optimization in Sec.~\ref{Subsec:Adaptive_Theta} can be directly extended to token-wise erasure channels by replacing the original angular distance $\Theta_{i,j}^{(m')}$ with its channel-aware counterpart $\tilde{\Theta}_{i,j}^{(m')}$. Equivalently, the inclusion indicator $A_{i,j}^{(m')}$ in \eqref{Eq:Theta_Obj}--\eqref{Eq:Theta_C1} is replaced by $\tilde{A}_{i,j}^{(m')} = \mathbb{I}\!\left(\tilde{\Theta}_{i,j}^{(m')} \le \theta_i\right)$, while the overall optimization structure remains unchanged.

   \section{Simulation Results}\label{Sec:Simul}
   In this section, we evaluate the effectiveness of the proposed geometry-based joint cross-modal token selection framework through simulations, with the following tasks:
    \begin{itemize}
        \item {\bf Visual question answering (VQA):}
        VQA is a task to generate the correct answer given an image and a natural language question about the image. 
        We use the VQA v2 dataset \cite{Goyal_VQAv2_2017}\footnote{VQA v2 is built on COCO images and contains 82,783 training images with 443,757 questions and 4,437,570 answers. Since the official test annotations are not publicly available, we use the validation set for testing, which includes 40,504 images, 214,354 questions, and 2,143,540 answers.}, where each sample consists of an image-question pair, and the model predicts the corresponding textual answer. We formulate the task as a multi-class classification problem over the top 1,000 most frequent answers in the training set, where the most frequent response among the 10 human annotations is used as the ground-truth label. The network is optimized using the cross-entropy loss, and performance is evaluated using top-1 classification accuracy.
        

        \item {\bf Audio-visual question answering (AVQA):}
        AVQA is a task to answer questions about visual objects, sounds, and their relationships in videos.
        We use the MUSIC-AVQA v2 \cite{Liu_MUSIC_AVQA_v2_2024} dataset\footnote{For training, 10,280 videos and 42,492 question-answer pairs are used. For evaluation, 7,071 videos and 10,747 question-answer pairs are used.}, where each sample consists of a video, its corresponding audio, and a natural language question, and the model predicts the corresponding textual answer. We formulate the task as a multi-class classification problem over 42 predefined answer classes. The network is optimized using the cross-entropy loss, and performance is evaluated using top-1 classification accuracy.

    \end{itemize}

    For both tasks, the proposed framework employs pretrained Transformer-based backbone encoders for modality-specific feature extraction. All visual inputs are resized to $(3,\!224,\!224)$ and normalized into tensor representations, while the corresponding natural language questions are tokenized using the standard BERT tokenizer \cite{Devlin_BERT_2019} and truncated to a maximum sequence length of $N_{\rm txt}=64$. The image encoder adopts the ViT-Base model \cite{ViT_model}, producing visual tokens with hidden dimension $D_{\rm img}=768$, and the text encoder adopts the BERT-Base model \cite{Devlin_BERT_2019}, producing text tokens with hidden dimension $D_{\rm txt}=768$. In VQA, a single image is used as input, yielding $N_{\rm img}=196$ visual tokens. In AVQA, each video is uniformly sampled into $8$ frames, resulting in $N_{\rm img}=196\times 8=1568$ visual tokens. In addition, the audio encoder adopts the Wav2Vec 2.0 Base model \cite{Baevski_Wav2Vec2_2020}, which processes a $16$ kHz mono waveform with a fixed duration of $10$ seconds and produces $N_{\rm aud}=1499$ audio tokens with hidden dimension $D_{\rm aud}=768$. We set $Q_m=32$ bits for all modalities.
    
    These modality-specific tokens are then processed by the proposed
    sequential cross-attention layer with $H=8$ and $d=96$, followed by
    feed-forward networks with a dropout rate of $0.1$. The modalities are
    updated in descending order of token count, allowing modalities with
    larger token sets to provide contextual information for subsequent
    updates. Accordingly, VQA adopts the update order image $\rightarrow$ text, whereas AVQA adopts the update order image $\rightarrow$ audio $\rightarrow$ text.
    For both tasks, the selected cross-modal tokens are linearly projected to a decoder dimension of $D_{\rm dec}=512$, concatenated with a learnable CLS token, and then processed by a 4-layer Transformer-based answer decoder with $H_{\rm dec}=8$ attention heads for final reasoning and classification. 
    
    The entire framework is fine-tuned offline using the cross-entropy loss. We adopt the AdamW optimizer \cite{Loshchilov_AdamW_2019} with weight decay $1\times10^{-4}$ and momentum parameters $\beta=(0.9,0.999)$. To effectively optimize the randomly initialized modules while preserving the robust representations learned by the pretrained encoders, differential learning rates are employed by assigning $1\times10^{-5}$ to the backbone encoders and $5\times10^{-5}$ to the cross-attention layer and answer decoder. The final prediction is formulated as a classification problem over 1,000 answer classes for VQA and 42 answer classes for AVQA.

    For both latency-constrained and token-wise erasure channels, we set $W=20~{\rm MHz}$, $\sigma^2=1$, and
    $\hat{R}_{\rm sum}=140~{\rm Mbps}$. For token-wise erasure channels, the bandwidth $W_\ell$ and transmit power $P_\ell^{(m)}$ allocated to each subchannel are adjusted based on Monte Carlo evaluations over Rayleigh fading realizations to achieve each prescribed $p_e^{(m)}$ according to the outage model in \eqref{eq:token_outage_probability}.

    \begin{figure*}[t]
    \begin{minipage}{2\columnwidth}
        \centering
        \subfigure[VQA, $M=2$.]
        {\epsfig{file=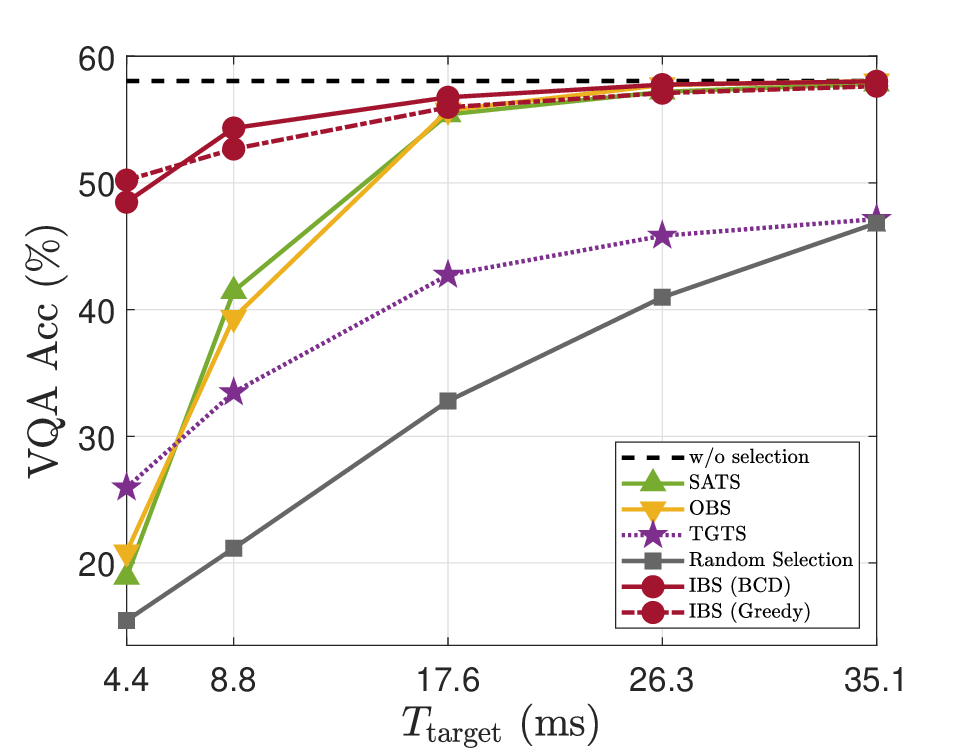, width=6.5cm}}
        \hspace{3mm}
        \subfigure[AVQA, $M=3$.]
        {\epsfig{file=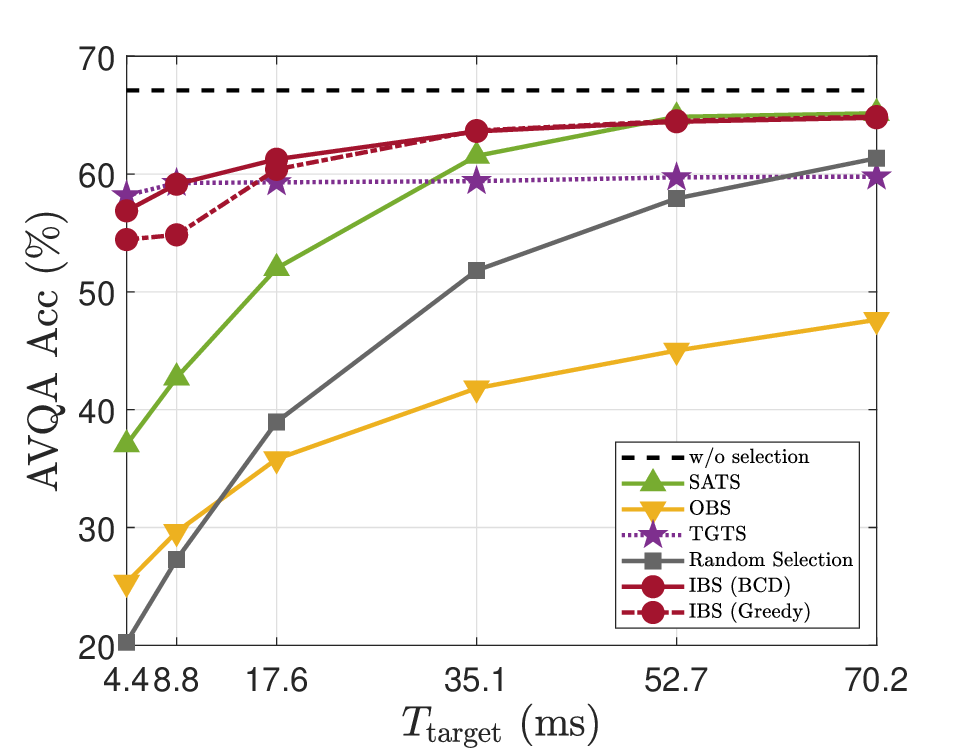, width=6.5cm}} 
        \captionof{figure}{
        Comparison of VQA and AVQA accuracy across various token selection schemes over latency-constrained wireless channels with $\hat{R}_{\rm sum}=140~{\rm Mbps}$.}
       \label{fig:vqa_acc_EF}
    \end{minipage}
    \end{figure*}

   For the VQA and AVQA tasks, we compare the proposed framework with the following baselines:
   \begin{itemize}
   
        \item {\bf IBS}: The proposed joint cross-modal token selection framework employs query-specific angular thresholds with an overlap criterion $k=2$. It is evaluated via three variants: {\bf IBS (BCD)} using the BCD-based solver (Sec.~\ref{Subsec:Adaptive_Theta_BCD}), {\bf IBS (Greedy)} using the low-complexity greedy algorithm (Sec.~\ref{Subsec:Adaptive_Theta_LC}), and {\bf R-IBS (Greedy)}, the channel-aware robust extension (Sec.~\ref{Sec:Channel_Aware_Metrics}).

        \item {\bf OBS} (Optimization-based selection): 
        This baseline directly maximizes the overall pairwise cross-modal relevance under the target latency. For each ordered modality pair $(m,m')$ with $m'\in\mathcal{M}\setminus\{m\}$, let ${\boldsymbol C}_h^{(m\to m')} \in \mathbb{R}^{N_m \times N_{m'}}$ denote the attention score matrix from the tokens of modality $m$ to those of modality $m'$ at the $h$-th attention head, obtained from the corresponding cross-attention update. For the two-modality case, this corresponds to the attention score matrices in \eqref{attn_score_i2t} and \eqref{attn_score_t2i}. Using the two directional matrices ${\boldsymbol C}_h^{(m\to m')}$ and ${\boldsymbol C}_h^{(m'\to m)}$, we construct the symmetrized cross-modal relevance matrix for the modality pair $(m,m')$ as
        \begin{align}
        {\boldsymbol C}_{m,m'}^{\mathrm{sym}}
        =
        \frac{1}{2H}
        \sum_{h=1}^{H}
        \left(\!
        {\boldsymbol C}_h^{(m \to m')}
        +
        \left({\boldsymbol C}_h^{(m' \to m)}\right)^{\!\! \top}
        \right),
        \end{align}
        for all $m,m' \in \mathcal{M}$ with $m < m'$. Using these pairwise relevance scores, OBS selects the cross-modal tokens with the largest aggregate relevance while satisfying the target latency $T_{\rm target}$.

        \item {\bf TGTS} (Text-guided token selection): Inspired by FastV \cite{Chen_FastV_2024}, this baseline adopts an asymmetric selection strategy guided by the text modality. It first treats text as the anchor modality and retains the text-side cross-modal tokens corresponding to the actual question tokens. The remaining target latency is then allocated to non-anchor modalities according to their relevance to the retained text anchors. For each non-anchor modality $m' \in \mathcal{M}\setminus\{{\rm txt}\}$, the relevance score of the $j$-th cross-modal token is defined as $\eta_j^{(m')}=\sum_{i=1}^{N_{{\rm txt}}} [{\boldsymbol C}_{{\rm txt},m'}^{\mathrm{sym}}]_{i,j}$. The remaining budget is assigned to the non-anchor tokens with the largest relevance scores across all $m' \in \mathcal{M}\setminus\{{\rm txt}\}$ while satisfying the latency constraint. Consequently, TGTS preserves the full question context while selecting non-anchor tokens in a text-guided manner.

        \item {\bf SATS} (Self-attention-based token selection): Inspired by EViT \cite{Liang_EViT_2022}, this baseline selects cross-modal tokens according to modality-wise importance scores derived from the self-attention matrices of the modality-specific encoders. To account for modality asymmetry, the target latency $T_{\rm target}$ is divided across modalities in a task-dependent manner. For the VQA task, $0.2T_{\rm target}$ is assigned to the text modality, and the remaining latency is assigned to the image modality. For the AVQA task, $0.03T_{\rm target}$ and $0.2T_{\rm target}$ are assigned to the text and audio modalities, respectively, and the remaining latency is assigned to the image modality. For a fair comparison, SATS is implemented as a post-hoc selection scheme at the encoder output, retaining cross-modal tokens with the largest self-attention-based importance scores within each modality until the target latency is reached.

        \item {\bf Random Selection}: This baseline uniformly selects cross-modal tokens from the joint candidate pool consisting of the  question-related text anchors and all non-anchor tokens from the remaining modalities, until the accumulated latency reaches $T_{\rm target}$.

   \end{itemize}

    \subsection{Performance Evaluation under Latency-Constrained Wireless Channels}

    \begin{figure*}[t]
        \begin{minipage}{2\columnwidth}
            \centering
            \subfigure[VQA ($M=2$, $p^{(\rm img)}_e = 0.2$).]
            {\epsfig{file=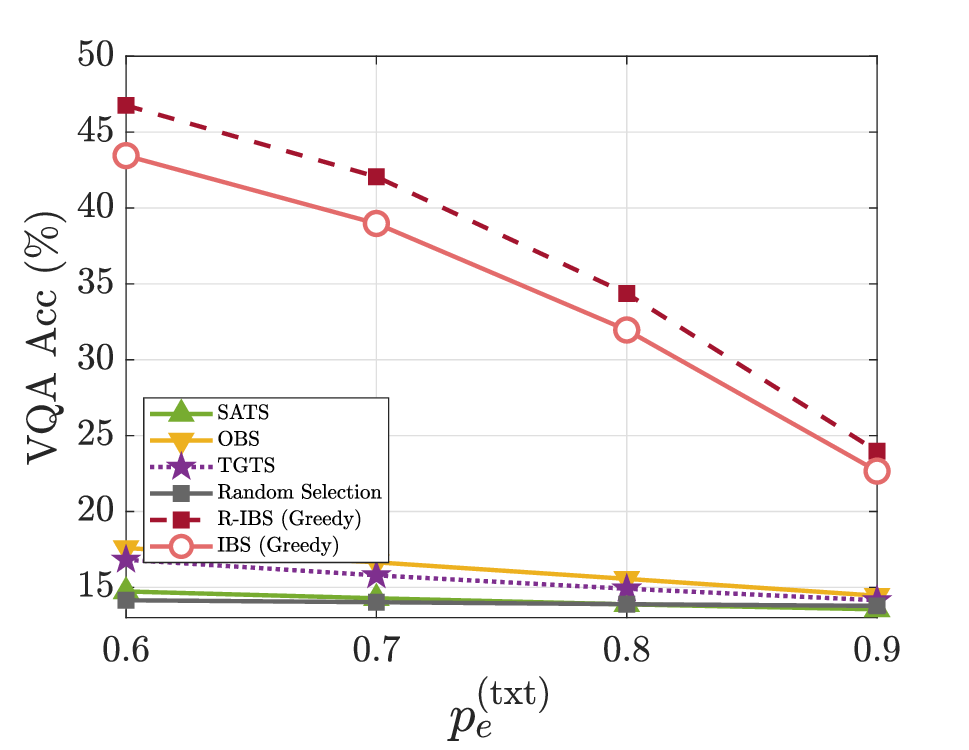, width=6.5cm}}
            \hspace{3mm}
            \subfigure[AVQA ($M=3$, $p^{(\rm img)}_e=p^{(\rm aud)}_e=p^{(\rm txt)}_e$).]
		{\epsfig{file=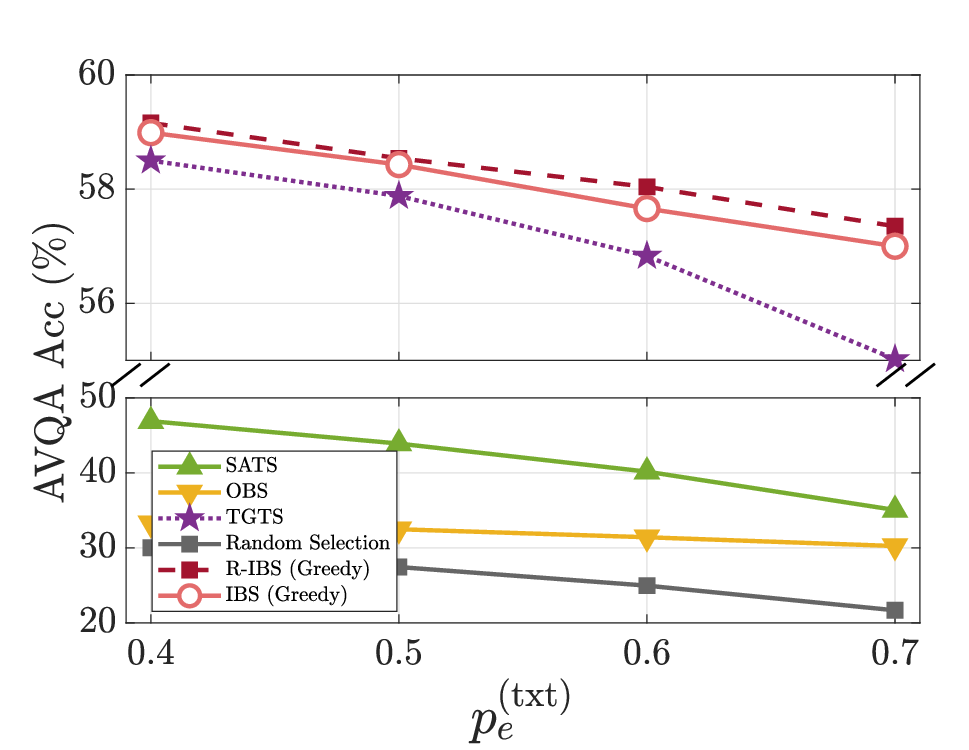, width=6.7cm}} 
            \captionof{figure}{
            Comparison of VQA and AVQA accuracy across various token selection schemes over token-wise erasure channels with different erasure probabilities under $\hat{R}_{\rm sum}=140~{\rm Mbps}$. The target latency is set to $T_{\rm target} = 4.4~{\rm ms}$ for VQA and $T_{\rm target} = 17.6~{\rm ms}$ for AVQA.}
           \label{fig:vqa_acc_ER}
        \end{minipage}
    \end{figure*}

    Fig.~\ref{fig:vqa_acc_EF} compares the accuracy of various token
    selection schemes on the VQA and AVQA tasks over latency-constrained wireless channels. In Fig.~\ref{fig:vqa_acc_EF}(a), {\bf IBS (BCD)} achieves the best overall performance across the entire $T_{\rm target}$ range and remains the closest to {\bf w/o selection}. {\bf IBS (Greedy)} also exhibits highly competitive performance and slightly outperforms {\bf IBS (BCD)} at $T_{\rm target} = 4.4~{\rm ms}$. As $T_{\rm target}$ increases, however, {\bf IBS (BCD)} becomes consistently superior, indicating that the BCD-based optimization more effectively exploits the enlarged feasible selection space available at higher $T_{\rm target}$ values.
    Both IBS variants consistently outperform {\bf Random}, supporting the
    geometric alignment hypothesis in Sec.~\ref{Subsec:Shared_Space} that
    higher cosine similarity improves downstream inference.

    {\bf OBS} and {\bf SATS} achieve competitive performance but remain
    inferior to the proposed IBS-based schemes. 
    At $T_{\rm target}=4.4~{\rm ms}$, {\bf IBS (BCD)} improves VQA accuracy over {\bf OBS} by $27.6\%$, supporting the multi-query consensus hypothesis in Sec.~\ref{Subsec:IBS} that pairwise relevance alone does not fully capture shared cross-modal semantics.
    The gain over {\bf SATS} further demonstrates the benefit of modeling
    cross-modal relationships rather than estimating token importance
    independently within each modality.
    
    A particularly large performance gap is observed between {\bf IBS (BCD)} and {\bf TGTS}, with {\bf IBS (BCD)} improving VQA accuracy by up to $22.5\%$ at $T_{\rm target} = 4.4~{\rm ms}$. {\bf TGTS} first preserves the text-side cross-modal tokens corresponding to the question and then allocates the remaining latency to the non-anchor modalities. Thus, its selection is primarily determined by a fixed modality-prioritized selection rule. By contrast, the proposed framework can additionally retain text-side cross-modal tokens that are not directly tied to the explicit question tokens but still encode globally aggregated textual context and cross-modal information through the encoder and cross-attention layers. As a result, the proposed framework preserves a richer multimodal context than {\bf TGTS}, which leads to a clear performance advantage. Similar overall tendencies are also observed in Fig.~\ref{fig:vqa_acc_EF}(b) for the AVQA task, where {\bf IBS (BCD)} again achieves the highest or competitive performance across the considered $T_{\rm target}$ range.

    \subsection{Performance Evaluation under Token-Wise Erasure Channels}


    \begin{figure}[t]
    \centering   
    {\epsfig{file=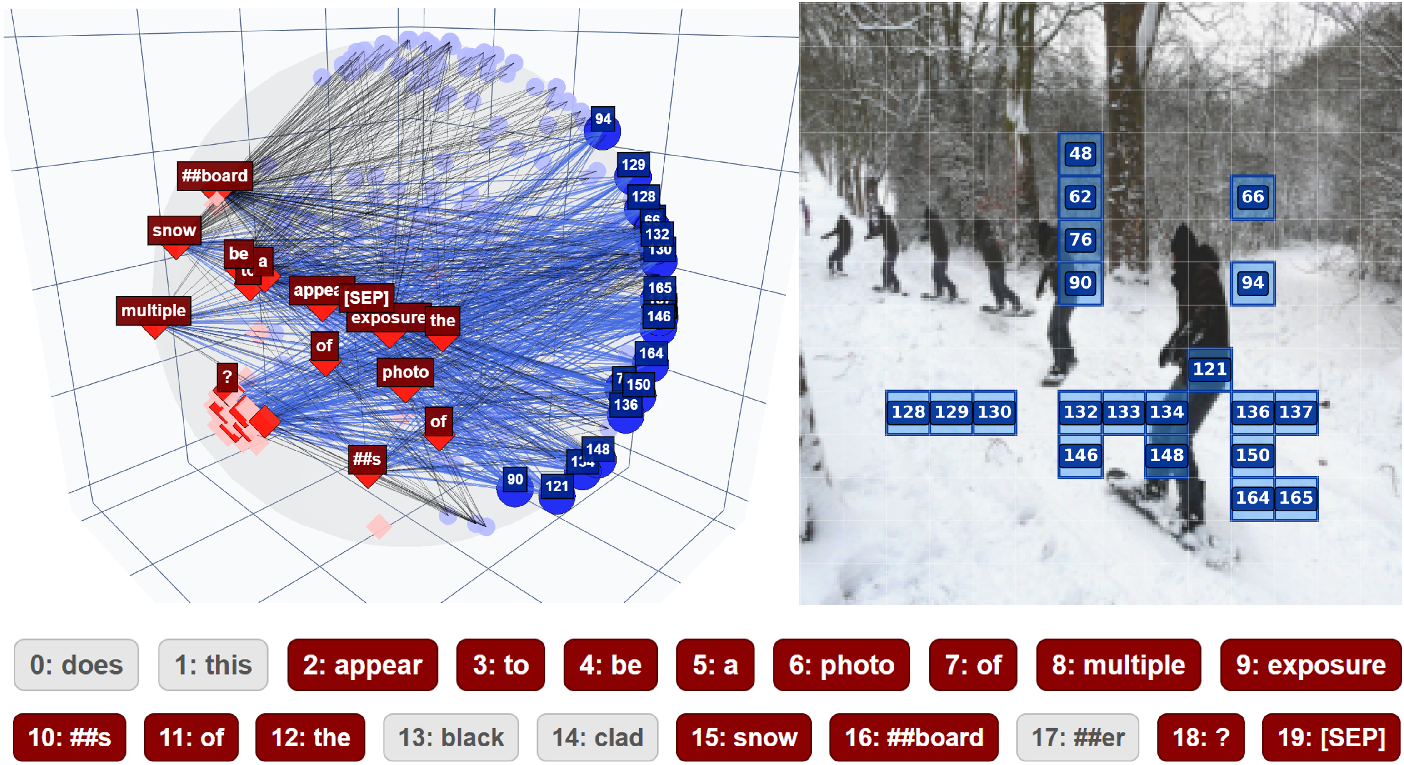, width=8.7cm}}
    \caption{
    Visualization of the cross-modal token selection result of {\bf IBS (BCD)} with $T_{\rm target} = 7~{\rm ms}$ and $\hat{R}_{\rm sum}=140~{\rm Mbps}$. The left panel shows the shared query-key space projected into a three-dimensional space via multidimensional scaling, while the right and bottom panels show the selected image patches and text tokens in the original data domain.}
    \label{fig:all_selection_results}
    \end{figure}

    Fig.~\ref{fig:vqa_acc_ER} evaluates the robustness of various token selection schemes over token-wise erasure channels. In Fig.~\ref{fig:vqa_acc_ER}(a), the VQA accuracy is compared under different erasure probabilities $(p_e^{({\rm img})},p_e^{({\rm txt})})$ with a target latency of $T_{\rm target}=4.4~{\rm ms}$. As the text erasure probability increases from $0.6$ to $0.9$, all schemes suffer performance degradation due to the increasing loss of transmitted semantic tokens. However, the proposed {\bf R-IBS (Greedy)}, which incorporates the channel-aware expected cosine similarity in \eqref{Eq:Expected_CS_general}, consistently outperforms its channel-agnostic counterpart {\bf IBS (Greedy)}. This demonstrates that explicitly accounting for token-wise erasure probabilities during cross-modal token selection improves robustness under unreliable wireless links.
    
    Compared with conventional baselines, the proposed IBS-based schemes maintain a clear performance advantage across all tested erasure conditions. In particular, at $(p_e^{({\rm img})},p_e^{({\rm txt})})=(0.2,0.6)$, {\bf R-IBS (Greedy)} achieves a $29.2\%$ accuracy gain over {\bf OBS}, the strongest conventional baseline in Fig.~\ref{fig:vqa_acc_ER}(a). Moreover, {\bf IBS (Greedy)} already outperforms {\bf SATS}, {\bf TGTS}, and {\bf Random Selection}, suggesting that the proposed IBS framework is effective even without channel-aware adaptation.

    Similar tendencies are observed in Fig.~\ref{fig:vqa_acc_ER}(b) for the AVQA task with $T_{\rm target}=17.6~{\rm ms}$. As the erasure probabilities increase across image, audio, and text modalities, {\bf R-IBS (Greedy)} again achieves the highest overall robustness, while {\bf IBS (Greedy)} remains superior to the conventional baselines. 

    Fig.~\ref{fig:all_selection_results} visualizes the cross-modal token selection result of {\bf IBS (BCD)} for the VQA task by projecting the shared query-key space into a three-dimensional space via multidimensional scaling \cite{Torgerson_MDS_1952}. 
    Since this projection approximately preserves relative distances, a higher cosine similarity between a text query and an image key tends to appear as a shorter angular distance in the projected space. In the three-dimensional visualization, red diamonds and blue circles denote selected text queries and image keys, respectively, while unselected tokens are shown as fainter semi-transparent markers. Solid blue lines indicate text-query--image-key pairs retained by the IBS rule, whereas gray lines denote candidate connections that lie within a semantic grain region but are discarded because the corresponding image keys do not satisfy the IBS overlap condition. The selected tokens are also mapped back to the original input for interpretation. 
    The highlighted image patches correspond to visually informative regions, including the snowboarder and the surrounding snowy motion regions, while the selected text tokens retain key concepts related to the VQA question, such as ``photo,'' ``multiple,'' ``exposure,'' and ``snowboard.'' 
    This result shows that {\bf IBS (BCD)} jointly selects semantically relevant text-side and image-side cross-modal tokens within the target latency by exploiting their query-key alignment in the shared space.

   \section{Conclusion}\label{Sec:Conclusion}
    In this paper, we proposed a geometry-based joint cross-modal token selection framework for latency-constrained multimodal token communications. 
    By mapping cross-modal tokens into a shared query-key space, we derived an angular-distance criterion for joint anchor and non-anchor token selection.
    Based on this, we developed the IBS strategy, which prioritizes the selection of non-anchor tokens whose key representations lie within the semantic grain regions of multiple anchor queries. The angular-distance thresholds defining these regions are optimized using BCD and a low-complexity greedy algorithm.
    Simulation results on a bimodal VQA task and a trimodal AVQA task demonstrated that the proposed framework consistently outperformed conventional baselines in terms of both task accuracy and channel robustness under latency-constrained and token-wise erasure channels.
    An important direction for future work is to extend the proposed framework to multi-user multimodal communication systems by jointly adapting cross-modal token selection and resource allocation to user-specific modality requirements, channel conditions, and task relevance.

\bibliographystyle{IEEEtran}
\bibliography{Reference_v2}
\end{document}